\documentclass[aps,pra,groupedaddress,amsmath,amssymb,galley,11pt,tightenlines,notitlepage,nofootinbib]{revtex4-1}

\usepackage[left=1.25in,right=1.25in,top=1in,bottom=1in]{geometry}
\usepackage{graphicx}
\usepackage[dvipsnames,svgnames]{xcolor}
\usepackage{bm,bbm}%
\usepackage{braket}
\usepackage{amsthm,amsmath,amssymb,amsbsy}
\usepackage{enumerate}
\usepackage{subcaption,caption}
\usepackage{booktabs,multirow}
\usepackage{mathtools,mathrsfs}
\usepackage[percent]{overpic}
\usepackage{microtype}

\usepackage[utf8]{inputenc}

\usepackage{array,adjustbox,afterpage}

\definecolor{darkred}{rgb}{0.57,0,0.12}
\usepackage[colorlinks=true, linkcolor=MidnightBlue, urlcolor=darkred!75!black, citecolor=MidnightBlue, pdfpagemode=UseOutlines]{hyperref}

\usepackage{cleveref,makecell}
\usepackage[most,breakable]{tcolorbox}

\AtBeginDocument{
\heavyrulewidth=.08em
\lightrulewidth=.05em
\cmidrulewidth=.03em
\belowrulesep=.65ex
\belowbottomsep=0pt
\aboverulesep=.4ex
\abovetopsep=0pt
\cmidrulesep=\doublerulesep
\cmidrulekern=.5em
\defaultaddspace=.5em
}

\makeatletter
\renewcommand{\p@subsection}{}
\renewcommand{\p@subsubsection}{}
\def\l@subsubsection#1#2{}
 \def\@hangfrom@section#1#2#3{\@hangfrom{#1#2}#3}%
 \def\@hangfroms@section#1#2{#1{#2}}%
 \def\@alph#1{\ifcase#1\or a\or b\or c\or d\else\@arabic{#1}\fi}
 \let\@fnsymbol@latex\@fnsymbol
 \let\@fnsymbol\@arabic
 \def\thempfootnote@latex{{\itshape \@arabic \c@mpfootnote }}%
 \def\ltx@thempfootnote{\@arabic\c@mpfootnote}
\makeatother

\usepackage{pxfonts,newpxmath}
\usepackage{etoolbox}
\hypersetup{linktoc=all}

\let\mathds\mathbbm

\DeclareMathOperator{\Tr}{Tr}
\DeclareMathOperator{\SR}{SR}
\DeclareMathOperator{\CR}{CR}

\DeclareMathOperator{\supp}{supp}
\DeclareMathOperator{\esupp}{esupp}
\DeclareMathOperator{\diag}{diag}
\DeclareMathOperator{\conv}{conv}
\DeclareMathOperator{\dom}{dom}
\DeclareMathOperator{\rank}{rank}
\DeclareMathOperator{\sspan}{span}
\DeclareMathOperator{\relint}{relint}
\let\Re\relax
\DeclareMathOperator{\Re}{Re}

\newcommand{\norm}[2]{\left\lVert#1\right\rVert_{\,#2}}
\newcommand{\proj}[1]{\ket{#1}\!\bra{#1}}
\newcommand{\ketbra}[2]{\ket{#1}\!\bra{#2}}

\newcommand{\lnorm}[2]{\left\lVert#1\right\rVert_{\ell_{#2}}}

\newcommand{\VV}{V^{\,\C}_\Sp}

\let\sect\S
\renewcommand{\S}{\mathcal{S}}
\newcommand{\Sp}{{\mathcal{S}_{+}}}
\newcommand{\Spc}{{\mathcal{S}^\circ_{+}}}
\newcommand{\Spd}{{\mathcal{S}_{+}\hspace{-0.85ex}\raisebox{0.2ex}{\*}}}
\newcommand{\Spdd}{{\mathcal{S}_{+}\hspace{-0.85ex}\raisebox{0.2ex}{\*\*}}}

\newcommand{\Spsym}{{\Sp \cup (-\Sp)}}
\newcommand{\Sperp}{{\S^\perp_+}}

\newcommand{\W}{\mathcal{W}}
\newcommand{\T}{\mathcal{T}}

\newcommand{\Scup}{{\mathcal{S}^{\cup}_+}}

\newcommand{\Ck}{{\mathcal{C}^{k}}}
\newcommand{\Ckp}{{\mathcal{C}^{k}_+}}
\newcommand{\C}{\mathcal{C}}

\newcommand{\K}{\mathcal{K}}

\newcommand{\V}{\mathcal{V}}
\newcommand{\X}{\mathcal{X}}
\newcommand{\Y}{\mathcal{Y}}
\renewcommand{\P}{\mathcal{P}}
\newcommand{\M}{\mathcal{M}}
\newcommand{\I}{\mathcal{I}}

\newcommand{\cleq}{\preceq}
\newcommand{\cgeq}{\succeq}

\newcommand{\ext}[1]{\operatorname{ext}\left(#1\right)}

\newcommand{\ppt}{\text{PPT}}
\newcommand{\pptp}{\text{PPT}_{+}}

\renewcommand{\*}{\textup{*}}
\newcommand{\<}{\left\langle}
\renewcommand{\>}{\right\rangle}

\renewcommand{\bar}{\;\rule{0pt}{9.5pt}\right|\;}

\newcommand{\lset}{\left\{\left.}
\newcommand{\rset}{\right\}}
\DeclareMathOperator{\interior}{int}

\newcommand{\A}{{\Gamma}}

\newcommand{\RR}{\mathbb{R}}
\newcommand{\CC}{\mathbb{C}}

\newcommand{\HH}{\mathbb{H}}
\newcommand{\DD}{\mathbb{D}}

\newcommand{\cbraket}[1]{\left|\braket{#1}\right|}

\newcommand{\RNG}{\mathcal{O}_{\text{RNG}}}
\newcommand{\RF}{\mathcal{O}_{\text{SRNG}}}
\newcommand{\F}{\mathcal{O}}

\newcommand{\locc}{\textsc{locc}}
\newcommand{\PPT}{\textsc{ppt}}
\newcommand{\GMN}{\textsc{rgmn}}
\newcommand{\cptp}{\textsc{cptp}}

\newcommand{\cc}{{\circ\circ}}

\newcommand{\texteq}[1]{\stackrel{\mathclap{\mbox{\text{\scriptsize #1}}}}{=}}

\newcommand{\Rs}[1]{R^{\smash{#1}}_{#1}}
\newcommand{\Rg}[1]{R^{\DD}_{#1}}

\newcommand{\wt}[1]{#1'}

\renewenvironment{boxed}[1]%
	{\expandafter\ifstrequal\expandafter{#1}{orange}{\begin{tcolorbox}[colback=orange!3,colframe=orange!15]}{\begin{tcolorbox}[colback=white,colframe=gray!10,breakable,enhanced]}}%
	{\end{tcolorbox}}

\theoremstyle{definition}
\newtheorem{theorem}{Theorem}

\newtheorem{proposition}[theorem]{Proposition}
\newtheorem{corollary}[theorem]{Corollary}
\newtheorem{definition}[theorem]{Definition}
\newtheorem{lemma}[theorem]{Lemma}
\newtheorem*{remark}{Remark}
\let\oldproofname\proofname
\renewcommand{\proofname}{\rm\bf{\oldproofname}}

\patchcmd{\section}
  {\centering}
  {\raggedright\Large}
  {}
  {}
\patchcmd{\subsection}
  {\centering}
  {\raggedright\large}
  {}
  {}
  \patchcmd{\subsubsection}
  {\centering}
  {\raggedright\normalsize\bfseries\textup}
  {}
  {}

\makeatletter
\def\l@f@section{%
 \addpenalty{\@secpenalty}%
 \addvspace{0.4em plus\p@}%
}%
\makeatother

\begin{document}

\title{\Large Convex geometry of quantum resource quantification}

\author{Bartosz Regula\vspace*{-5pt}}
\affiliation{School of Mathematical Sciences, University of Nottingham,\\University Park, Nottingham NG7 2RD, United Kingdom}
\email{bartosz.regula@gmail.com}

\begin{abstract}
\noindent\makebox[\textwidth][c]{%
    \begin{minipage}{.9\textwidth}
\vspace{10pt}
We introduce a framework unifying the mathematical characterisation of different measures of general quantum resources and allowing for a systematic way to define a variety of faithful quantifiers for any given convex quantum resource theory. The approach allows us to describe many commonly used measures such as matrix norm--based quantifiers, robustness measures, convex roof--based measures, and witness-based quantifiers together in a common formalism based on the convex geometry of the underlying sets of resource-free states. We establish easily verifiable criteria for a measure to possess desirable properties such as faithfulness and strong monotonicity under relevant free operations, and show that many quantifiers obtained in this framework indeed satisfy them for any considered quantum resource. We derive various bounds and relations between the measures, generalising and providing significantly simplified proofs of results found in the resource theories of quantum entanglement and coherence. We also prove that the quantification of resources in this framework simplifies for pure states, allowing us to obtain more easily computable forms of the considered measures, and show that several of them are in fact equal on pure states. Further, we investigate the dual formulation of resource quantifiers, characterising the dual sets of resource witnesses.
\vspace{10pt}\\
We present an explicit application of the results to the resource theories of multi-level coherence, entanglement of Schmidt number $k$, multipartite entanglement, as well as magic states, providing insight into the quantification of the four resources by establishing novel quantitative relations and introducing new quantifiers, such as a measure of entanglement of Schmidt number $k$ which generalises the convex roof--extended negativity, a measure of $k$-coherence which generalises the $\ell_1$ norm of coherence, and a hierarchy of norm-based quantifiers of $k$-partite entanglement generalising the greatest cross norm.
 \end{minipage}}
\end{abstract}

\maketitle

{\vspace*{10pt}
 \footnotesize\tableofcontents}

\section{Introduction}

Many physical phenomena in quantum information science have gone from being of purely theoretical interest to enjoying a variety of uses as \textit{resources} in quantum information processing tasks. The developments sparked an investigation into the mathematical formulation of such resource theories, aiming to characterise the quantum states and operations that one can use to perform the physical tasks. Starting with the resource theory of entanglement, which found use in a wide variety of quantum information processing, quantum communication, and quantum computing protocols \cite{horodecki_2009}, the recent years have seen the establishment of resource theories of athermality \cite{brandao_2013}, asymmetry \cite{gour_2008,lostaglio_2015}, purity \cite{horodecki_2003}, coherence \cite{aberg_2006,gour_2008,baumgratz_2014,streltsov_2017}, nonclassicality \cite{vogel_2014,theurer_2017}, EPR steering \cite{gallego_2015}, contextuality \cite{grudka_2014}, magic states \cite{veitch_2014,ahmadi_2017}, and others, including more general mathematical formulations of resource theories \cite{horodecki_2012,sperling_2015,brandao_2015,delrio_2015,coecke_2016,gour_2017,liu_2017}.

In particular, it is crucial to be able to \textit{quantify} the given resource, allowing us to discriminate which quantum states are the most useful in the given physical task. Throughout the development of the resource theory of entanglement, various measures were established \cite{plenio_2007,horodecki_2009}, many of which have been adapted to other resource theories recently \cite{baumgratz_2014,brandao_2015,piani_2016,howard_2017,streltsov_2017,gour_2017,theurer_2017}. However, defining and characterising the measures of a given quantum resource is usually cumbersome --- the investigation of such functions typically has to be approached in a resource-dependent way, and properties such as faithfulness and monotonicity of the quantifiers have to be explicitly verified. Moreover, although some connections between the various quantities are known, there are very few known results which provide a common framework relating them and their features together. 

\subsection{Summary of the results}

In this work, we introduce a unifying formalism based on the convex geometry of the underlying sets of quantum states, significantly simplifying the construction and characterisation of quantifiers of general quantum resources. We employ the concept of gauge functions, a fundamental tool in functional and convex analysis \cite{rudin_1991,rockafellar_1970,chandrasekaran_2012}, to establish a consolidated view of many resource measures. In particular, we show that many commonly used and well-known quantifiers --- such as ones based on matrix norms, measures built through the convex roof, the so-called robustness measures, as well as various witness-based quantifiers --- are all examples of such gauge functions, allowing us to relate them in a common geometric framework. This lets us establish an extensive family of quantifiers for any given quantum resource, introduce easily verifiable criteria for a measure to satisfy desirable properties such as faithfulness and strong monotonicity under relevant free operations, and generalise known measures to new quantum resources very easily. Further, we show that many relations and bounds between the measures, some of which known in the resource theories of entanglement and coherence, are in fact universal among quantum resources, and the proofs of such properties can be simplified in the present framework.

The formalism of this paper applies to general finite-dimensional resource theories with a convex set of resource-free states, which is a common and intuitive assumption \cite{sperling_2015,brandao_2015}. A particularly useful case of such resources, and one that we will focus on, is when the set of free states is obtained as the convex hull of free pure states. More specifically, given a set of pure state vectors $\V$ (these can be e.g. product pure states in the resource theory of entanglement, or the reference basis vectors in the resource theory of coherence), we define the set of interest --- the free states $\Sp$ --- as the convex hull of projectors $\proj\psi$ with each $\ket\psi \in \V$. The quantities of interest are then the \textit{gauge functions} corresponding to different sets, which can be understood as an alternative notion of ``distance'' from the given set. Gauge functions have a number of appealing properties such as a rich structure of convex duality, allowing many quantifiers to admit a simplified characterisation when expressed in this way.

The simplest and easiest to compute gauge function is $\A_\V$, that is, the gauge based on the set $\V$. This gauge can be used to quantify pure-state resources --- for instance, in the resource theory of bipartite entanglement, it is equal to the sum of Schmidt coefficients \cite{vidal_2002}, while in the resource theory of coherence it corresponds to the $\ell_1$ norm \cite{baumgratz_2014}. The quantification of resources for general mixed states has a more complex structure and one can establish a variety of different gauges which all form valid quantifiers of the given resource. In fact, we can show that many well-known monotones belong to the gauge function formalism: these are the robustness measures, generalising the fundamental entanglement measures of robustness $\Rs\Sp$ \cite{vidal_1999} and global robustness $\Rg\Sp$ \cite{steiner_2003,piani_2016,brandao_2015}; the norm-based measures $\A_\S$, which include the greatest cross norm of entanglement \cite{rudolph_2000,rudolph_2001} and the $\ell_1$ norm of coherence \cite{baumgratz_2014}; the convex roof--based measures $\A^\cup_\Sp$, such as the convex roof--extended negativity \cite{lee_2003} and the coherence concurrence \cite{yuan_2015,qi_2017}; as well as several other measures, such as many experimentally-friendly families of witness-based quantifiers known from the theory of entanglement \cite{brandao_2005,guhne_2007-1,eisert_2007}.

The crucial application of the framework is that the often extremely technical and cumbersome resource-dependent proofs of properties, bounds, and analytical expressions for the resource measures are not needed, as the convex geometric framework provides simplified proof methods and establishes relations which hold regardless of the considered resource.

\enlargethispage{1\baselineskip} To begin with, the gauge function formalism helps establish a fundamental property of the above quantities: for any considered quantum resource, all of the gauge-based measures are in fact valid resource quantifiers, satisfying desirable properties such as faithfulness, convexity, and strong monotonicity under relevant classes of operations. Further, the framework provides easily verifiable conditions for any other gauge-based quantity to share the same properties, applicable e.g. to various witness-based measures.

We then establish several quantitative relations between the gauges for any given convex quantum resource, which immediately allow us to relate the introduced quantifiers with each other, generalising results from resource theories of bipartite entanglement and coherence to general quantum resources. Furthermore, in the convex geometric framework each gauge function has an associated dual gauge function, allowing us to establish several bounds and equivalences which can be useful in the characterisation of the dual sets of resource witnesses. We additionaly relate the gauges with other fundamental quantifiers such as the distance- and witness-based measures, showing that more connections can be generalised to arbitrary convex resource theories.

We further show that the bounds obtained above can in fact be tight: for any convex resource theory, several of the considered quantifiers reduce to the simplified gauge $\A_\V$ on pure states:
\begin{equation}
  \A_\S(\proj\psi) = \Rg\Sp(\proj\psi)+1 = \A^\cup_\Sp(\proj\psi) = \A_\V(\ket\psi)^2.
\end{equation}
These relations, frequently a non-trivial fact to establish for a particular resource theory, show that the quantification of pure-state resources is always simplified. Note that the gauge function $\A_\V$ is often significantly easier to compute that the general forms of the quantifiers, in many cases leading to an analytical characterisation of pure-state resource measures.

To exemplify the application and usefulness of the framework, we consider some representative resource theories --- bipartite and multipartite entanglement, quantum coherence, and magic states --- obtaining new results in the quantification of the resources, and in particular novel bounds and analytical formulas for quantifiers of bipartite entanglement of Schmidt rank $k$, $k$-partite entanglement, and $k$-coherence. In addition to shedding new light on quantifiers already defined in the literature, we introduce several new measures, such as: a measure of multi-level quantum coherence which generalises the $\ell_1$ norm of coherence, faithful quantifiers of magic, measures of bipartite entanglement of Schmidt number $k$ and $k$-partite entanglement which generalise the convex roof--extended negativity, as well as a class of norms which generalise the greatest cross norm to the hierarchy of $k$-partite entanglement, with computable formulas for genuine multipartite entanglement. 

The paper is structured as follows. In section \ref{sec:gauges}, we review methods from convex geometry and analysis, and provide an introduction to the concept of gauge functions as well as the convex roof. In section \ref{sec:quantum} we apply the framework to quantum states, showing how to define measures for any convex resource theory and establishing results concerning the computability and interrelations between many such quantities. Section \ref{subsec:resource_quantifiers} contains a characterisation of atomic gauges as quantifiers of a given quantum resource, establishing easily verifiable conditions for properties such as faithfulness and strong monotonicity, as well as relating the gauge functions to measures defined through resource witnesses and distance-based quantifiers. Finally, section \ref{sec:app} contains an explicit application of our results to several quantum resource theories as mentioned above.

\section{Gauges and norms}\label{sec:gauges}

\subsection{Atomic gauges}

The definitions in this section follow standard literature, in particular Rockafellar \cite[\sect 14-15]{rockafellar_1970}; see also \cite{hiriart-urruty_1993,bach_2013,friedlander_2014}, where much of this information is reviewed and expanded on. We will use $\RR_+$ to denote non-negative reals and $\RR_{++}$ to denote positive reals. A set $\K$ is called a cone if it contains any non-negative scalar multiple of its elements, i.e. $c \in \K \Rightarrow \lambda c \in \K\; \forall \lambda \in \RR_+$. A convex cone $\K$ is then a cone which contains any conic combination (non-negative linear combination) of elements in $\K$.

Consider a finite-dimensional real vector space with an inner product $\< \cdot, \cdot \>$. Given a set $\S$, we define its \textbf{polar set} $\S^\circ$ as
\begin{equation}
  \S^\circ = \lset x \bar \<x, s\> \leq 1\; \forall s \in \S \rset
  \end{equation}
and its \textbf{dual cone} $\S\*$ as
\begin{equation}
  \S\* = \lset x \bar \<x, s\> \geq 0\; \forall s \in \S \rset.
\end{equation}
The polar set and dual cone are always closed and convex, regardless of whether $\S$ is. The bipolar set, given by
\begin{equation}\begin{aligned}
  \S^\cc \coloneqq (\S^\circ)^\circ = \overline{\conv\left(\S \cup \{0\}\right)}
\end{aligned}\end{equation}
where $\overline{\vphantom{a}\,\cdot\,}$ denotes closure, is the smallest closed convex set containing $\S$ and the origin, and analogously the conic hull (bidual cone) $\S\*\*$ is the smallest closed convex cone containing $\S$. Note that the polar set of a cone $\K$ is given by $\K^\circ = -\K\*$.

The \textbf{gauge function} $\gamma_\S$ for a set $\S$ (also known as the Minkowski functional) is defined as
\begin{equation}
  \gamma_\S(x) = \inf \lset\lambda \in \RR_+ \bar x \in \lambda \S\rset = \inf_{s \in \S} \lset\lambda \in \RR_{+} \bar x = \lambda s \rset
\end{equation}
where we note that the effective domain of $\gamma_\S$ is the cone generated by $\S$, that is $\lset \mu s \bar \mu \in \RR_+,\, s \in \S \rset \eqqcolon \RR_+ \S$, and we follow the convention that $\gamma_\S(x) = \infty\; \forall x \notin \dom(\gamma_\S)$, or equivalently $\inf \emptyset = \infty$.

An important example of gauge functions are norms, which are the gauge functions of their unit balls. Norms are defined as finite functions satisfying the following axioms: absolute homogeneity (of degree 1), subadditivity, and positivity everywhere except the origin. This corresponds to gauge functions of sets $\S$ which are convex, compact, centrally symmetric ($\S=-\S$), and such that $0 \in \interior(\S)$. In general, we have the following relations between sets and their corresponding gauges \cite{rockafellar_1970,freund_1987}:
\begin{center}
\begin{tabular}{@{} *2l @{}}  \toprule
if $\S$ is convex and & then $\gamma_{\S}$ is \\\midrule
--- & \makecell[cl]{convex\\[-3pt]positively homogeneous, i.e. $\gamma_\S(\alpha x)\! =\! \alpha \gamma_\S(x)\forall \alpha\!\in\!\RR_{\!+\!+}$}\\
closed & lower semicontinuous \\
compact & positive everywhere except $0$, i.e. $\gamma_\S(x) > 0\;\forall x \neq 0$\\
centrally symmetric $\quad$ & symmetric (even), i.e. $\gamma_\S(x) = \gamma_\S(-x)$\\
$0 \in \interior(\S)$ & finite everywhere, i.e. $\gamma_\S(x) < \infty \; \forall x$\\\bottomrule
\end{tabular}
\end{center}
where we note that evenness together with positive homogeneity implies absolute homogeneity, and convexity with positive homogeneity implies subadditivity.

An important property of any gauge corresponding to a closed convex set $\S$ is that if $0 \in \S$, then the unit ball of $\gamma_\S$ is given exactly by $\S$, and each such unit ball \textit{uniquely} determines its corresponding gauge --- more generally, all closed convex gauge functions are in a one-to-one correspondence with closed convex sets containing the origin. In full generality, the unit ball of a convex gauge is given by
\begin{equation}
  \lset x \bar \gamma_\S(x) \leq 1 \rset = \S^{\circ\circ}
\end{equation}
which in particular means that $\gamma_\S = \gamma_{\S^{\circ\circ}}$.

Following the terminology of \cite{chandrasekaran_2012} and subsequent works, given a compact set $\C$ we define the \emph{atomic gauge} as the gauge function of the convex hull $\conv(\C)$:
\begin{equation}
  \A_\C(x) \coloneqq \gamma_{\conv(\C)} (x).
\end{equation}
The terminology comes from understanding $\C$ as a set of ``atoms'', often enjoying some simple structure which we want $\Gamma_\C$ to characterise. In particular, notice that $\Gamma_\C$ can be expressed as an optimisation over conic combinations of such atoms:
\begin{equation}\begin{aligned}\label{eq:atomic_norm}
  \A_\C(x) &= \inf \left\{ \lambda \;\left|\; x = \lambda \sum_i c_i a_i,\,a_i\in\C,\,c_i \in \RR_+,\,\sum_i c_i = 1  \right.\right\}\\
  &= \inf \lset \sum_i c_i \bar x = \sum_i c_i a_i,\,a_i\in\C,\,c_i \in \RR_+ \rset.
\end{aligned}\end{equation}
Recall that all norms are gauge functions, but in addition many common norms can be conveniently expressed as an optimisation over simple sets of atoms --- as two representative examples in $\RR^{d\times d}$, consider the trace norm (Schatten $1$-norm), which is given for $\C = \lset x y^T \bar x,y \in \RR^d,\; \lnorm{x}{2} = \lnorm{y}{2} = 1 \rset$, and the operator norm (Schatten $\infty$-norm), which corresponds to $\C$ being the set of all orthogonal matrices. We will look into the structure of similar norms in the sequel, where a suitable choice of atoms will enable us to develop a general geometric description of quantum resources.

By Carath\'{e}odory's theorem for cones, it suffices to consider combinations of at most $d$ elements in \eqref{eq:atomic_norm}, with $d$ denoting the dimension of the vector space.  Note that in general $\Gamma_\C$ need not be symmetric or finite everywhere, and so it only defines a valid norm when the linear span of $\C$ is the whole space and $\C$ is centrally symmetric. The domain of $\A_\C$ is the conic hull of $\C$, given by $\RR_+\conv(\C) = \C\*\*$. Notice that $\conv(\C)^\circ = \C^\circ$ and $\conv(\C)\* = \C\*$.

The \textbf{polar function} of $\A_\C$ is defined as
\begin{equation}\begin{aligned}
  \A_\C^\circ(x) &= \sup \lset \<x,a\> \bar \A_\C(a) \leq 1 \rset\\
  &= \smashoperator{\sup_{a \in \C^{\circ\circ}}}\;\<x,a\>\\
  &= \max\left\{ 0,\,\sup_{a \in \C}\;  \<x,a\>\right\}
\end{aligned}\end{equation}
where the last equality follows from the fact that the supremum of a linear functional over a compact convex set is reached at an extremal point of the set. The polar of the atomic gauge is precisely the atomic gauge for the polar set $\C^\circ$, i.e. $\A^\circ_\C = \A_{\C^\circ}$, and in fact $\A_{\C^\circ} = \gamma_{\C^\circ}$ as $\C^\circ$ is already a closed convex set containing the origin ($\C^{\circ\circ\circ} = \C^\circ$). The polarity operation induces a one-to-one symmetric correspondence between closed convex gauge functions, which means that any atomic gauge of a compact set is uniquely determined by its polar. When $\A_\C$ defines a valid norm, $\A_\C^\circ$ is its dual norm.

A closed set $\C$ is bounded if and only if $0 \in \interior(\C^\circ)$; dually, $\C^\circ$ is bounded if and only if $0 \in \interior(\C^{\circ\circ})$. For any bounded set $\C$, $\A^\circ_\C$ then has full domain. Note that the polarity operation is inclusion-reversing, i.e. $\C_1 \subseteq \C_2 \Rightarrow \C^\circ_1 \supseteq \C^\circ_2$; this means that $\A_{\C_1}(x) \geq \A_{\C_2}(x) \; \forall x\in\dom(\A_{\C_1})$ and $\A^\circ_{\C_1}(y) \leq \A^\circ_{\C_2}(y)\;\forall y\in\dom(\A_{\C_2}^\circ)$ for such sets.

Since $0 \in \C^\circ$ for any set $\C$, by considering the polar function of the set $\C^\circ$ we can also write
\begin{equation}
  \A_\C(x) = \A_{\C^{\circ\circ}}(x) = \sup_{a \in \C^\circ}\; \<x,a\>
\end{equation}
which we will refer to as the dual formulation of the atomic gauge. We note that all atomic gauges also satisfy some properties which are well-known for norms, such as the generalised Cauchy-Schwarz inequality $\<x, y\> \leq \A_\C(x) \,\A_\C^\circ(y) \; \forall x\in\dom(\A_\C),\,y\in\dom(\A_\C^\circ)$.

We will hereafter mostly work with atomic gauges, and for simplicity we will often write $\A_\C$ even if $\C$ is already convex and the notation is superfluous.

\subsubsection{Complex vector spaces}

In a complex vector space, we define the polar with respect to the real inner product $\Re\<x,y\>$, and similarly for other definitions. A generalisation of the concept of a centrally symmetric set is then a \textit{balanced set}, that is $\C$ s.t. $s \in \C \Rightarrow \xi s \in \C\;\forall \xi\in\CC :|\xi|=1$. In particular, for a balanced set we have $\A_\C(x) = \A_\C(\xi x)$ for all $\xi$ as above, and we can then take a simplified definition of a polar function:
\begin{equation}
  \A_\C^\circ(x) = \max\left\{ 0,\, \sup_{a \in \C}\; \Re\<x,a\> \right\} = \sup_{a \in \C}\;\left|\<x,a\>\right|.
\end{equation}
Any convex, balanced, compact set which spans the whole space defines a valid norm. We will henceforth only encounter balanced sets in complex vector spaces, so we do not expand on other properties and definitions --- see e.g. \cite{bourbaki_2003} for more general cases.

\subsection{Convex roof}
The intuition behind the convex roof approach is due to Uhlmann \cite{uhlmann_1998,uhlmann_2010}. He describes the basic idea of the procedure as a way to extend entanglement measures ``as linearly as possible'' to the set of mixed states. It is formalised as follows.

Let $\S$ be a compact convex set, and let $\ext\S$ be the set of its extreme points, such that $\S = \conv(\ext\S)$. Given a function $f$ defined on $\ext\S$, we now would like to consider a function $\wt{f}$ defined on the whole set $\S$ such that the two functions are equal on $\ext\S$ --- we will call such functions extensions of $f$. Extensions of this kind were briefly investigated in the mathematical literature before Uhlmann's work \cite{lima_1972,peters_1986} and have also received attention in other areas after finding use in quantum information \cite{tawarmalani_2002,bucicovschi_2013,yan_2014,shcherbatyi_2016}. A particularly useful class of them is defined as follows:

\setcounter{theorem}{1}
\begin{definition} A function $\wt{f}: \S \rightarrow \RR_+$ is a \textbf{roof extension} of $f: \ext\S \rightarrow \RR_+$ if for every point $x \in \S$  there exists at least one extremal convex decomposition $\{c_i, \pi_i\}$ of the form $x = \sum_i c_i \pi_i$ with $c_i \in \RR_+$, $\sum_i c_i = 1$, $\pi_i \in \ext\S$, such that $\wt{f} (x) = \sum_i c_i\, f(\pi_i)$.
\end{definition}
The justification for the terminology of \textit{roofs} can be understood by noting that if we know a set of extremal points $\{\pi_1, \ldots, \pi_k\}$ which constitutes an optimal decomposition for some $x \in \S$, the roof extension $\wt{f}$ will be affine on the convex hull of these points --- that is, the same set of points will be an optimal decomposition for any other convex combination thereof. This means that the graph of $\wt{f}$ consists of flat (affine) pieces covering the set $\S$, not unlike a roof covering a floor.

This definition, however, might seem arbitrary --- since there are no restrictions on the choice of a roof, it does not tell us much about the properties of the function $f$. Here is where the convexity of $\wt{f}$ comes useful.
\begin{boxed}{white}
\begin{theorem}[Uhlmann \cite{uhlmann_1998}]\label{thm:uhlmann} If a convex roof extension $f^\cup$ of $f: \ext\S \rightarrow \RR_+$ onto $\S$ exists, then it is unique and satisfies the following properties:
\begin{enumerate}[(i)]
\item $f^\cup$ is the pointwise largest convex extension of $f$ from $\ext\S$ to $\S$
\item $f^\cup$ is the pointwise smallest roof extension of $f$ from $\ext\S$ to $\S$
\item $f^\cup$ is given by
\begin{equation}\label{eq:def_conv_roof}
f^\cup(x) = \inf \lset \sum_i c_i\, f(\pi_i) \bar \sum_i c_i \, \pi_i = x,\; \sum_i c_i = 1,\; c_i \in \RR_+ \rset
\end{equation}
with the infimum taken over all extremal convex decompositions $\{c_i, \pi_i\}$ with each $\pi_i \in \ext\S$.
\end{enumerate}
\end{theorem}
\end{boxed}
The result tells us that finding the largest convex function on $\S$ which coincides with $f$ on $\ext\S$ is equivalent to finding the unique convex roof extension of $f$. For a lower semicontinuous function $f$, an optimal decomposition realising the infimum in eq. \eqref{eq:def_conv_roof} always exists.

Since we will be dealing with functions $f: \ext\S \to \RR_+\cup\{\infty\}$, we extend the definition by taking $f^\cup(x) = \infty$ when $x \notin \conv(\dom f)$. For a function $g$ already defined on the whole set $\S$, we can define its convex roof extension as the convex roof extension of the restriction of $g$ to $\ext\S$ --- if $g$ is convex, it is then clear from the above Theorem that $g^\cup(x) \geq g(x)\,\forall x$.

Note also the related concept of the \textit{concave roof} $f^\cap$: it is the pointwise smallest concave extension (or pointwise largest roof extension) of $f$, and can be calculated by replacing the minimisation with a maximisation in Eq. \eqref{eq:def_conv_roof}:
\begin{equation}
  f^\cap(x) = \sup_{\{c_i, \pi_i\}} \lset \sum_i c_i\, f(\pi_i) \bar \sum_i c_i \, \pi_i = x,\; \sum_i c_i = 1,\; c_i \in \RR_+ \rset.
\end{equation}

One can notice a similarity between the concept of the convex roof and the \textit{convex envelope} (also known as the \textit{convex hull} or the \textit{largest convex underestimator}). For a function $g$ which is already defined on the whole convex set $\S$, its convex envelope is the largest convex function smaller than $g$. In practice, finding the convex envelope means that the optimisation in eq. \eqref{eq:def_conv_roof} is performed over all decompositions of $x$ into $s_i \in \S$, instead of $\pi_i \in \ext\S$ as in the case of the convex roof extension. For this reason, in quantum information literature the convex envelope has sometimes been referred to as the \textit{mixed convex roof}, identifying $\ext\S$ with the set of pure quantum states. Note that the convex envelope of a concave function defined on $\S$ coincides with its convex roof extension, since the infimum of any concave function over a bounded convex set is equal to the infimum over its extremal points \cite[32.2]{rockafellar_1970}.

A simple relation between the convex roof and the formalism of atomic gauges can be obtained as follows.
\begin{boxed}{orange}
\begin{proposition}\label{prop:any_gauge_conv_roof}Given a compact set $\C \subseteq \ext\S$, for any $x \in \S$ we have
\begin{equation}
   \A_\C(x) = \gamma^\cup_{\C}(x).
 \end{equation}
 \end{proposition}
 \end{boxed}
\begin{proof}
The atomic gauge $\Gamma_\C$ can be written as
\begin{align}
  \A_\C(x) &= \inf \lset \sum_i c_i \bar x = \sum_i c_i x_i ,\; c_i \in \RR_+,\; x_i \in \C \rset\nonumber\\
  &\texteq{(i)} \inf \lset \sum_i c_i \,\gamma_\C(x_i) \bar x = \sum_i c_i x_i ,\; c_i \in \RR_+,\; x_i \in \C \rset\\
  &\texteq{(ii)} \inf \lset \sum_i c_i \,\gamma_\C(x_i) \bar x = \sum_i c_i x_i,\; c_i \in \RR_+,\; \sum_i c_i = 1,\; x_i \in \C\*\*\rset \nonumber\\
  &\texteq{(iii)} \inf \lset \sum_i c_i \,\gamma_\C(x_i) \bar x = \sum_i c_i x_i,\; c_i \in \RR_+,\; \sum_i c_i = 1,\; x_i \in \ext\S\*\*\rset\nonumber\\
  &= \gamma^\cup_{\C}(x)\nonumber.
\end{align}
The step (i) follows since, without loss of generality, we can consider only $x_i$ such that $\gamma_\C(x_i) = 1$. To see this, notice that for any $c'_i x'_i$ with $\gamma_\C(x'_i) \in (0,1)$ we can define $x_i \coloneqq x'_i / \gamma_\C(x'_i)$ and $c_i \coloneqq c'_i \gamma_\C(x'_i)$ such that $c'_i \gamma_\C(x'_i) = c_i \gamma_\C(x_i) = c_i$. The step (ii) follows similarly, and (iii) is a straightforward consequence of the fact that $\gamma_\C(x_i) = \infty$ for any $x_i \notin \C\*\*$, so the optimisation is effectively constrained to be over $x_i \in \C\*\*$, in the sense that the infimum (if finite) will necessarily be given for $x$ in a decomposition into $x_i \in \C\*\*$.
\end{proof}
We remark that, following a very similar reasoning as in the proof above, one can write any gauge function of this form as
\begin{equation}\begin{aligned}
  \A_\C(x) &= \inf \lset \sum_i \gamma_\C(x_i) \bar x = \sum_i x_i,\; x_i \in \ext\S \rset.\\
\end{aligned}\end{equation}
The applications of this result and further relations of this kind will be explored in subsequent chapters.

The application of Proposition \ref{prop:any_gauge_conv_roof} is particularly useful for sets of positive semidefinite matrices, where we identify $\S$ with $\HH_+$ and $\ext\S$ with rank-one positive semidefinite matrices. As an explicit example, given a balanced set $\X \subseteq \CC^d$, let us define $\C \subseteq \ext{\HH_+}$ as
\begin{equation}
\C = \lset \proj{a} \bar \ket{a} \in \X \rset.
\end{equation}
Then, one can easily see that $\gamma_\C(\proj{x}) = \gamma_\X(\ket{x})^2$, which allows us to express the atomic gauge function of $\C$ as
\begin{equation}
  \A_\C(P) = \gamma^\cup_\C(P) = \inf \lset \sum_i p_i \gamma_\X(\ket{x_i})^2 \bar P = \sum_i p_i \proj{x_i},\, \sum_i p_i = 1,\,p_i\in \RR_+ \rset.
\end{equation}
We will extend this idea by considering other convex roof extensions for sets of matrices in the succeeding sections.

\section{Atomic gauges for sets of quantum states}\label{sec:quantum}

One of the essential elements of the characterisation of a general resource theory are the \textbf{free states}, that is, states not possessing a given resource. We therefore begin by defining the set of resource-free normalised pure state vectors $\V \subseteq \CC^d$, which we will assume to be non-empty. Another intuitive assumption is that the set $\V$ should be compact, which ensures the continuity of the given resource theory \cite{brandao_2015}. Since the global phase factor $e^{i\theta}\ket{\psi}$ of a quantum state $\ket\psi$ is physically irrelevant, we further assume that $\ket\psi \in \V$ implies $e^{i\theta}\ket\psi \in \V \;\forall\theta\in\RR_+$, which means the set is balanced and so $0 \in \conv(\V)$. We then get the atomic gauge
\begin{align}
  \A_\V(\ket\psi) &= \inf \lset \sum_i c_i \bar \ket\psi = \sum_i c_i \ket{v_i},\,\ket{v_i} \in \V,\, c_i \in \RR_+ \rset\label{eq:gaugeV_definition1}\\
  &= \!\!\sup_{\ket x \in \V^\circ} \cbraket{\psi|x}\nonumber\\
  \A_\V^\circ(\ket\psi) &= \!\sup_{\ket y \in \V} \cbraket{\psi|y}.\label{eq:gaugeV_definition2}
\end{align}
Under the assumptions above, $\A_\V$ is a valid norm for $\CC^d$ whenever $0 \in \interior(\conv\V)$, i.e. $\sspan(\V) = \CC^d$. This property is very desirable for the set $\V$, as we will see later, but we do not assume it to hold in all cases. Note that, by the compactness of $\V$, the infima and suprema in \eqref{eq:gaugeV_definition1}-\eqref{eq:gaugeV_definition2} will be achieved as long as they are finite.

Given the set $\V$ defined at the level of vectors, we now define our set of interest: the set of resource-free density matrices, given by the convex hull of the set
\begin{equation}
\S_+ \coloneqq \lset \proj{v} \bar \ket{v} \in \V \rset \subseteq \DD.
\end{equation}
Since all density matrices which do not lie in $\conv(\Sp)$ are resourceful states in the given resource theory, we would now like to use the set $\Sp$ to introduce a gauge function which could effectively quantify this resource. The atomic gauge $\A_\Sp$ itself is not very useful: since the domain of this function is $\Spdd = \RR_+ \conv(\Sp)$ (the convex cone generated by $\Sp$) and all density operators lie in the hyperplane defined by $\<\mathbbm{1},\rho\>=1$, there are actually no density operators in the domain of $\A_\Sp$ which do not lie in the convex hull of $\Sp$. That is, for any density matrix $\rho$ we have
\begin{equation}\begin{aligned}
  &\A_\Sp(\rho) \leq 1 && \text{if } \rho \in \conv(\Sp),\\
  &\A_\Sp(\rho) = \infty && \text{if } \rho \notin \conv(\Sp).
\end{aligned}\end{equation}
We will now look at ways of circumventing this problem by making a different choice of the gauge function to use.


\subsection{Selection of gauges}\label{sect:basic}

Throughout the rest of this manuscript, we will be working in several different vector spaces: the underlying Hilbert space, identified with $\CC^d$; the real vector space of self-adjoint linear operators, idenitifed with the space of Hermitian matrices $\HH$ and endowed with the real inner product $\<A,B\>=\Tr(AB)$; as well as the complex vector space of linear operators acting on $\CC^d$, identified with the space of complex square matrices $\CC^{d \times d}$ and endowed with the (degenerate) real inner product $\Re \<A,B\> = \Re \Tr(A^\dagger B)$. Unless otherwise specified, the default setting will be the real vector space $\HH$.

We will use $\HH_1$ to denote the set $\lset X \in \HH \bar \<\mathbbm{1}, X\> = 1 \rset$ of unit trace Hermitian matrices, $\HH_+$ to denote the cone of positive semidefinite matrices, and $\DD = \HH_1 \cap \HH_+$ to denote the set of valid density matrices. We will use $\ket{x}$ to denote vectors in the Hilbert space which are not necessarily normalised, and reserve the standard Greek letters $\psi,\phi$ etc. to represent normalised pure states. Similarly, Greek letters $\rho,\sigma,\omega$ will denote normalised density operators.

Our aim now is to define a choice of non-trivial atomic gauges which provide a natural way to quantify the given resource.

An intuition for the possible choices of suitable quantifiers can be obtained by looking at the simple example where $\V$ is taken to be the set of \textit{all} normalised pure state vectors, and thus $\Gamma_\V(\cdot)= \lnorm{\cdot}{2}$. In this case, we have $\conv(\Sp) = \DD$, and the gauge $\A_\DD$ can be thought of as the trace norm $\norm{\cdot}{1}$, but defined only on the limited domain $\RR_+\DD = \HH_+$. It is straightforward to see that the convex roof extension of $\Gamma_\V$ corresponds to the same function. A natural way to extend the domain of this gauge beyond $\HH_+$ is to symmetrise the set under consideration, and consider the gauge $\A_{\DD \cup (-\DD)}$ instead --- since the origin is now contained in the convex hull of the set, the function takes finite values for all Hermitian matrices. In fact, $\A_{\DD \cup (-\DD)}$ is simply equal to the trace norm itself \cite{hartkamper_1974}, with the domain now extended to $\HH$. One can go a step further and, instead of limiting ourselves to the real vector space of Hermitian matrices, define the set $\S = \lset \ketbra{a}{b} \bar \ket{a}, \ket{b} \in \V \rset$. The gauge $\A_\S$ is then precisely the trace norm $\norm{\cdot}{1}$ in its most general formulation, which has full domain in $\CC^{d \times d}$.

In these simple examples, we have seen that different ways of extending the domain of the gauge $\A_\Sp$ actually correspond to the same function (on their effective domains), and so there is little ambiguity in the choice of the appropriate extension. This does not, however, hold in more general cases, and so it becomes useful to be able to characterise all the different possible gauges that we can obtain from the initial set $\V$.

Returning to the case of general quantum resources, we can take inspiration from the above procedure and consider the atomic gauge with respect to the symmetrised set $\Sp \cup (-\Sp)$. This extends the domain of the gauge function to $\sspan(\Sp)$, that is, the vector space generated by $\Sp$. We then get the gauge
\begin{equation}\begin{aligned}\label{eq:atomic_base}
\A_{\Sp \cup (-\Sp)}(\rho) &= \;\enskip\inf_{\mathclap{S_i \in \Spsym}}\quad\;\, \lset \sum_i c_i \bar \rho = \sum_i c_i\, S_i,\; c_i \in \RR_+ \rset \\
&= \inf_{\ket{v_i} \in \V} \lset \sum_i |c_i| \bar \rho = \sum_i c_i\, \proj{v_i},\; c_i \in \RR \rset\\
\A^\circ_{\Sp \cup (-\Sp)}(\rho) &= \sup \lset \cbraket{v|\rho|v} \bar \ket{v} \in \V \rset.
\end{aligned}
\end{equation}
Analogously to Prop. \ref{prop:any_gauge_conv_roof}, the above can be equivalently expressed as
\begin{equation}\begin{aligned}\label{eq:atomic_base_alt}
  \A_{\Sp \cup (-\Sp)}(\rho) &= \inf_{X_i \in \HH} \lset \sum_i \A_{\Sp \cup (-\Sp)}(X_i) \bar \rho = \sum_i X_i,\; \rank(X_i) = 1 \rset\\
  &= \!\inf_{\ket{x_i} \in \CC^d}\! \lset \sum_i |c_i|\, \A_{\Spsym}\!\left(\proj{x_i}\right) \bar \rho = \sum_i c_i \proj{x_i},\; c_i \in \RR \rset\!.
\end{aligned}\end{equation}
$\A_\Spsym$ defines a valid norm for Hermitian matrices as long as $\sspan(\Sp) = \HH$.

Another straightforward way to define a quantifier is to consider instead the set
\begin{equation}\S \coloneqq \lset\ket{a}\!\bra{b} \bar \ket{a} \in \V,\, \ket{b} \in \V\rset,\end{equation}
which we note to be balanced in the complex vector space $\CC^{d \times d}$ with the Hilbert-Schmidt inner product $\Re \< \cdot, \cdot \>$. The corresponding atomic gauge is:
\begin{equation}\begin{aligned}
\A_\S(\rho) &= \inf_{\ket{v_i}, \ket{w_i} \in \V} \lset \sum_i c_i \bar \rho = \sum_i c_i\, \ket{v_i}\!\bra{w_i},\; c_i \in \RR_+ \rset\\
\A_\S^\circ(\rho) &= \sup \lset \cbraket{v|\rho|w} \bar \ket{v},\ket{w} \in \V \rset.
\end{aligned}
\end{equation}
This gauge admits an alternative representation, as we show below. The result is a slight generalisation of a lemma in \cite{richard_2014,richard_2014-1}, where norms for real matrices are considered.
\setcounter{theorem}{0}
\begin{boxed}{orange}
\begin{proposition}\label{lemma:nuclear}
$\displaystyle\A_\S(\rho) = \inf_{\ket{x_i}, \ket{y_i} \in \CC^d} \lset \sum_i \A_\V(\ket{x_i})\,\A_\V(\ket{y_i}) \bar \rho = \sum_i \ket{x_i}\!\bra{y_i} \rset.$
\end{proposition}
\end{boxed}
\begin{proof}
The function on the right-hand side can be noticed to be the gauge function of the set
\begin{equation}\begin{aligned}
  \S' \coloneqq& \lset \ket{a}\!\bra{b} \bar \A_\V(\ket{a}) \leq 1,\, \A_\V(\ket{b}) \leq 1 \rset\\
   =& \lset \ket{a}\!\bra{b} \bar \ket{a}, \ket{b} \in \V^\cc \rset.
 \end{aligned}\end{equation}
The straightforward inclusion $\S \subseteq \S'$ gives $\S^\cc \subseteq \S'{}^\cc = \S'$, as $\S'$ is already a closed convex set containing the origin. On the other hand, noticing that for any $\ket{a}\!\bra{b} \in \S'$ we can write $\ket{a} = \sum_i c_i \ket{a_i}$ and $\ket{b} = \sum_j d_j \ket{b_j}$, where $c_i, d_j \in \RR_+$, $\ket{a_i}, \ket{b_j} \in \V$, and $\sum_i c_i,  \sum_j d_j \leq 1$, we get $\ket{a}\!\bra{b} = \sum_{i,j} c_i d_j \ket{a_i}\!\bra{b_j}$ with $\sum_{i,j} c_i d_j \leq 1$, which means that $\ket{a}\!\bra{b} \in \S^\cc$ and so $\S' \subseteq \S^\cc$. The sets $\S'$ and $\S^\cc$ are therefore equal, and so $\Gamma_{\S} = \Gamma_{\S'}$.
\end{proof}
\setcounter{theorem}{4}
\noindent Gauges of the form encountered on the right-hand side of Prop. \ref{lemma:nuclear} are referred to as nuclear gauges \cite{jameson_1987,bach_2013}. $\A_\S$ defines a valid norm for $\CC^{d\times d}$, called a generalised nuclear norm, as long as $\A_\V$ is a norm for $\CC^d$, i.e. $\sspan(\V) = \CC^d$.

Considering the values of the gauge function $\A_\V$ for pure states is appealing, as this function is typically significantly simpler to compute than gauges defined at the level of matrices. We can then extend $\Gamma_\V$ to general quantum states through the convex roof. To ensure homogeneity at the level of projectors $\proj\psi$, we will take the extension of $\A_\V(\ket\psi)^2$, giving
\begin{equation}\begin{aligned}
  \A^\cup_\Sp(\rho) \coloneqq& \left(\A^2_\V\right)^\cup (\rho)\\
  =& \inf \lset \sum_i p_i \A_\V(\ket{\psi_i})^2 \bar \rho = \sum_i p_i \proj{\psi_i},\, \sum_i p_i = 1,\,p_i\in \RR_+ \rset
\end{aligned}\end{equation}
where the optimisation is over normalised pure states. The convex roof extension can therefore be regarded as a natural extension of $\A_\Sp$ out of its limited domain, since it will be finite for any density matrix as long as $\sspan(\V) = \CC^d$. We can characterise it as a gauge function as follows.


\begin{boxed}{orange}
\begin{proposition}\label{prop:convex_roof_gauge}
For any $\rho \in \DD$, the convex roof extension $\A^\cup_\Sp(\rho)$ is equal to the atomic gauge $\Gamma_{\Scup}(\rho)$, where
\begin{equation}
  \Scup \coloneqq \lset \proj{a} \bar \ket{a} \in \conv(\V) \rset.
\end{equation}
\end{proposition}
\end{boxed}
\begin{proof}
By Prop. \ref{prop:any_gauge_conv_roof}, the gauge function of $\Scup$ can be written as
\begin{equation}\begin{aligned}
  \Gamma_{\Scup}(\rho) = \inf \lset \sum_i p_i \A_\V(\ket{x_i})^2 \bar \rho = \sum_i p_i \proj{x_i},\, \sum_i p_i = 1,\,p_i\in \RR_+ \rset,
\end{aligned}\end{equation}
where we now optimise over general rank-one terms. The result follows simply by noting that we can renormalise any such decomposition $\rho = \sum_i p_i \proj{x_i}$ as $\rho = \sum_i p'_i \proj{\psi_i}$ where $p'_i \coloneqq p_i \lnorm{\ket{x_i}}{2}^2$ and $\ket{\psi_i} \coloneqq \ket{x_i}/\lnorm{\ket{x_i}}{2}$, giving
\begin{equation}\begin{aligned}
  \sum_i p'_i = \sum_i p_i \Tr \proj{x_i} = \Tr\left( \sum_i p_i \proj{x_i}\right) = \Tr \rho = 1.
\end{aligned}\end{equation}
Using the homogeneity of $\Gamma_\V$, this ensures that $\Gamma_{\Scup}(\rho) = \A^\cup_\Sp(\rho)$.
\end{proof}

Before we proceed with the characterisation of the introduced atomic gauges, let us make explicit the effective domains on which the functions are finite.
\begin{center}
\begin{tabular}{@{} *3l @{}}  \toprule
Gauge & Effective domain \hspace{4pt} & $\DD$ in the effective domain?\\\midrule
$\A_\V$ & $\sspan(\V)$ & N/A\\
$\A^\circ_\V$ & $\CC^d$ & N/A\\
$\A_\Sp$ & $\Spdd$ & no, $\dom(\A_\Sp) \cap \DD = \conv(\Sp)$\\
$\A_\Spsym$ & $\sspan(\Sp)$ & if $\relint_\DD(\conv \Sp) \neq \emptyset$\\
$\A_\S$ & $\sspan(\S)$ & if $\sspan(\V) = \CC^d$\\
$\A^\cup_\Sp$ & $\Scup\*\*$ & if $\sspan(\V) = \CC^d$\\
$\A^\circ_\Sp,\A^\circ_\Spsym,\A^{\cup\circ}_\Sp$ \hspace*{5pt}& $\HH$& yes\\
$\A^\circ_\S$ & $\CC^{d \times d}$& yes\\\bottomrule
\end{tabular}
\end{center}
In the above, $\relint_\DD(\conv \Sp)$ is the interior of $\conv(\Sp)$ relative to $\DD$. The set $\Scup\*\*$ can be understood as the set of positive semidefinite matrices supported on the subspace spanned by $\V$, i.e.
\begin{equation}\begin{aligned}
  \Scup\*\* = \lset X \in \HH_+ \bar \supp(X) \subseteq \sspan(\V) \rset.
\end{aligned}\end{equation}

We also remark that in general we have
\begin{equation}\begin{aligned}\label{eq:gauge_ineq}
  &\A_\Sp(\rho) \geq \A_\Spsym(\rho) \geq \A_\S(\rho)\\
  &\A_\Sp(\rho) \geq \A^\cup_\Sp(\rho) \geq \A_\S(\rho)
\end{aligned}\end{equation}
and reverse inequalities for the polar gauges, which follows from the set inclusion of their corresponding unit balls.


\begin{boxed}{orange}
\begin{theorem}\label{prop:nucl_pure_states}
For pure states $\proj\psi$ s.t. $\ket\psi \in \sspan(\V)$, we have
\begin{equation}\begin{aligned}
  \A^\cup_\Sp(\proj\psi) &= \A_\V(\ket\psi)^2\\
  \A_\S(\proj\psi) &= \A_\V(\ket\psi)^2.
\end{aligned}\end{equation}
For arbitrary pure states we also have
\begin{equation}\begin{aligned}
  \A^\circ_\Sp(\proj\psi) &= \A^\circ_\V(\ket\psi)^2\\
  \A_\S^\circ(\proj\psi) &= \A_\V^\circ(\ket\psi)^2.\\
  \end{aligned}
\end{equation}
\end{theorem}
\end{boxed}
\begin{proof}
We have shown the case of $\A^\cup_\Sp$ in the proof of Prop. \ref{prop:convex_roof_gauge}.

For $\A_\S$, notice that by the compactness of $\V$ there exist $\lambda^\star \in \RR_+$ and $\ket{v^\star} \in \conv(\V)$ such that
\begin{equation}
  \lambda^\star = \min \lset \lambda \in \RR_+ \bar \ket\psi \in \lambda\conv(\V) \rset = \A_\V(\ket\psi)
\end{equation}
with $\ket\psi = \lambda^\star \ket{v^\star}$. This gives $\proj\psi = \lambda^{\star 2} \proj{v^\star}$ with $\proj{v^\star} \in \conv(\S)$, and so
\begin{equation}
  \A_\V(\ket\psi)^2 = \lambda^{\star 2} \geq \inf \lset \eta \in \RR_+ \bar \proj\psi \in \eta \conv(\S) \rset = \A_\S (\proj\psi).
\end{equation}

Now, from the dual formulation of the atomic gauge we get
\begin{equation}
\begin{aligned}
  \A_\S (\proj\psi) &= \sup \lset \left|\< \proj\psi, X \>\right| \bar X \in \S^\circ \rset\\
  &\geq \sup \lset \left|\<\proj\psi, \proj{x} \>\right| \bar \ket{x} \in \V^\circ \rset\\
  &= \sup \left\{\left. \cbraket{\psi|x}^2 \;\right|\; \ket{x} \in \V^\circ \right\}\\
  &= \A_\V(\ket\psi)^2
  \end{aligned}
\end{equation}
since $\ket{x} \in \V^\circ \Rightarrow \proj{x} \in \S^\circ$, and so the claim follows.

The case of $\A_\Sp^\circ$ follows immediately from the definition. For $\A_\S^\circ$, we have
\begin{equation}
\begin{aligned}
  \A_\S^\circ(\proj\psi) &= \sup \lset \left\lvert \braket{x |\psi}\braket{\psi | y}\right\rvert \bar \ket{x}, \ket{y} \in \V \rset\\
  &= \left(\sup_{\ket{x} \in \V} \cbraket{x | \psi}\right)\left(\sup_{\ket{y} \in \V} \cbraket{y | \psi}\right)\\
  &= \A_\V^\circ (\ket\psi) \A_\V^\circ (\ket\psi)
  \end{aligned}
\end{equation}
since the optimisation is over two compact sets of non-negative real numbers.
\end{proof}
We can also note a stronger property that
\begin{equation}\begin{aligned}
  \A_\S (\ket\psi\!\bra\phi) &= \A_\V(\ket\psi)\, \A_\V(\ket\phi)\\
  \A_\S^\circ (\ket\psi\!\bra\phi) &= \A_\V^\circ (\ket\psi)\, \A_\V^\circ (\ket\phi).
\end{aligned}\end{equation}


\begin{boxed}{orange}
\begin{proposition}\label{prop:gauge_polars_equals}
For positive semidefinite matrices $P$, we have
\begin{equation}
  \A^\circ_\Sp(P) = \A^\circ_\S(P) = \A^{\circ}_\Scup(P) = \A^\circ_\Spsym(P)
\end{equation}
\end{proposition}
\end{boxed}
In the proof of the Proposition, we will make use of the following lemma.
\begin{boxed}{white}
\begin{lemma}\label{lemma:posdef}
  For any $P \in \HH_+$, it holds that
\begin{equation}
  \sup \lset \cbraket{a | P | b} \bar \ket{a},\ket{b}\in\V \rset = \sup \lset \braket{a | P | a} \bar \ket{a} \in \V \rset.
\end{equation}
\end{lemma}
\end{boxed}
\begin{proof}First, notice that the left-hand side clearly cannot be smaller than the right-hand side. By the compactness of $\V$, there exists a $\ket{a^\star}$ which realises the maximum on the right-hand side. Writing $P$ in its spectral decomposition $P = \sum_i \chi_i \proj{p_i}$, we then have for any $\ket{a}$, $\ket{b}$ that
\begin{equation}
  \begin{aligned}
  \cbraket{a|P|b}^2 &= \left| \sum_i \chi_i \braket{a | p_i}\braket{p_i | b} \right|^2\\
  &\leq \left(\sum_i \chi_i \cbraket{a | p_i}^2\right)\left(\sum_i \chi_i \cbraket{b | p_i}^2\right)\\
  &\leq \left(\sum_i \chi_i \cbraket{a^\star | p_i}^2\right)\left(\sum_i \chi_i \cbraket{a^\star | p_i}^2\right)\\
  &= \braket{a^\star | P | a^\star}^2
  \end{aligned}
\end{equation}
using the Cauchy-Schwarz inequality, which proves the Lemma.
\end{proof}
\begin{proof}\textit{(Proposition \ref{prop:gauge_polars_equals})}. By the above Lemma, one immediately obtains $\A^\circ_\Sp(P) = \A^\circ_\S(P)$ from the definitions. It is also straightforward to see that $\A^\circ_\Spsym(P)$ follows in a similar way. For $\A^{\circ}_\Scup$, we have that
\begin{equation}\begin{aligned}
  \A^{\circ}_\Scup (P) &= \sup \lset \< P, Z \> \bar Z \in \conv(\Scup) \rset\\
  &= \sup \lset \< P, Z \> \bar Z \in \Scup \rset\\
  &= \sup \lset \braket{\nu | P | \nu} \bar \ket{\nu} \in \conv(\V) \rset\\
  &= \sup \lset \braket{\nu | P | \nu} \bar \ket{\nu} \in \V \rset\\
  &= \sup \lset \<P, Z \> \bar Z \in \Sp \rset\\
  &= \A^\circ_\Sp(P)
\end{aligned}\end{equation}
where the fourth equality follows since each $\braket{\nu | P | \nu}$ is a convex quadratic form in $\ket{\nu}$ by the positive semidefiniteness of $P$ \cite{boyd_2004}, and the maximum of a convex function over a bounded convex set is reached at an extremal point of the set \cite[32.2]{rockafellar_1970}.
\end{proof}


\subsection{Base norms and robustness}\label{sec:robustness}

If the set $\Sp$ spans the whole space of Hermitian matrices, the gauge $\A_\Spsym$ defines a valid norm for $\HH$ called the \textit{base norm} \cite{hartkamper_1974}. Base norms have found a variety of uses in state and channel discrimination \cite{matthews_2009,reeb_2011,jencova_2014,lami_2017}. More generally, one can consider other gauges of the type $\A_{\Sp\cup(-\X)}$ with $\X$ being another set satisfying some desired properties. A connection between such gauges and measures commonly used in entanglement theory --- robustness and negativity --- was noticed by Vidal and Werner \cite{vidal_2002} as well as Rudolph \cite{rudolph_2005} and later briefly expanded on by Plenio and Virmani \cite{plenio_2007}. We extend and further generalise the relation between gauges of this type and resource quantifiers.

Consider a general quantity $R^\X_\Sp$, which we will refer to as the \textit{robustness} with respect to the closed set $\X \subseteq \DD$:
\begin{equation}\begin{aligned}
R^\X_\Sp(\rho) &= \enskip\inf_{\mathclap{\omega \in \conv(\X)}}\quad \lset s \in \RR_+ \bar \rho + s\omega \in \Spdd \rset\\
&= \enskip\inf_{\mathclap{\sigma \in \conv(\Sp)}}\quad \lset s \in \RR_+ \bar \rho - (1+s)\,\sigma \in (-\X)\*\* \rset\\
\end{aligned}\end{equation}
where the second equality follows by noting that $\rho \in \DD$ together with $\X \subseteq \DD$ mean that $\rho = r \sigma - s \omega \,\Rightarrow\, r = (1+s)$.
The robustness can be understood as the smallest amount of mixing with a state in the set $\conv(\X)$ necessary in order for the resulting renormalised mixed state to be in $\conv(\Sp)$ \cite{vidal_1999}. In the case that no feasible $s \in \RR_+$ exists, we take $\inf \emptyset = \infty$ as before.

In terms of gauge functions, the robustness can be expressed as the gauge corresponding to an unbounded set of the form:
\begin{equation}\begin{aligned}
	R^\X_\Sp(\rho) &= \A_{{\S_{+}\hspace{-0.85ex}\raisebox{0.05ex}{\footnotesize\*\*}}\cup(-\X)}(\rho)\\
	&= \A_{\Sp\cup(-\X)\*\*}(\rho) - 1.\\
\end{aligned}\end{equation}
Alternatively, by introducing the generalised inequality $A \cgeq_{\X} B \iff A - B \in \X\*\*$, we can also write the above as
\begin{equation}\begin{aligned}
R^\X_\Sp(\rho) &= \enskip\inf_{\mathclap{\omega \in \conv(\X)}}\quad \lset s \in \RR_+ \bar \rho \cgeq_\Sp -s\, \omega \rset\\
&=\enskip\inf_{\mathclap{\sigma \in \conv(\Sp)}}\quad \lset s \in \RR_+ \bar \rho \cleq_\X (1+s)\, \sigma \rset.
\end{aligned}\end{equation}
We note the similarity of this expression to geometric concepts such as the Thompson metric \cite{thompson_1963,cobzas_2014} and Hilbert's metric \cite{hilbert_1895,bushell_1973}, the latter having been applied to the study of distinguishability norms in quantum information \cite{reeb_2011}.

Comparing this with the expression for $\A_{\Sp\cup(-\X)}$, which we can write as
\begin{equation}
  \A_{\Sp\cup(-\X)}(\rho) = \inf \lset \lambda^+ + \lambda^- \bar \rho = \lambda^+ \sigma - \lambda^- \omega,\;\lambda^\pm \in \RR_+,\;\sigma \in \conv(\Sp),\; \omega \in \conv(\X) \rset,
 \end{equation}
we have that $\rho \in \DD \Rightarrow \lambda^+ - \lambda^- = 1$ and so the following relation holds:
\begin{equation}
  R^\X_\Sp(\rho) = \frac{\A_{\Sp\cup(-\X)} - 1}{2}.
\end{equation}

Two choices of the set $\X$ will be particularly useful: following the naming conventions of the theory of quantum entanglement, we introduce the \textbf{(standard) robustness} $\Rs\Sp$ and the \textbf{generalised robustness} $\Rg\Sp$ \cite{vidal_1999}. The latter is, up to logarithm, equivalent to the so-called max-relative entropy \cite{datta_2009}. We will hereafter omit the subscript in $\cgeq_\DD$ and simply use $\cgeq$ to refer to the inequality with respect to the positive semidefinite cone (L\"{o}wner partial order).

By the above consideration, the standard robustness is an affine function of the base gauge $\A_\Spsym$, and the generalised robustness of $\A_{\Sp\cup(-\DD)}$. We recall that the effective domain of $\A_\Spsym$ is $\sspan(\Sp)$, and so again we have $\DD \subseteq \dom(\Rs\Sp)$ only if $\relint_\DD(\conv \Sp) \neq \emptyset$. The domain of the generalised robustness $\Rg\Sp$ consists of all states supported on the subspace spanned by $\V$, i.e. $\dom(\Rg\Sp) = \Scup\*\*$, and so $\sspan(\V) = \CC^d$ is necessary and sufficient for $\Rg\Sp$ to be finite for all states.

We now consider the dual formulations of the gauges. Noting that $\left(\Sp \cup -\DD\right)^\circ = \Spc \cap (-\DD)^\circ$ and similarly for $\Spsym$, we have
\begin{align}\label{eq:gen_robustness_dual_gauge}
&\Rg\Sp(\rho) = \frac{1}{2} \sup \lset \<\rho, W'\> \bar W' \in \Spc \cap (-\DD)^\circ \rset - \frac{1}{2}\\
\label{eq:standard_robustness_dual_gauge}
&\Rs\Sp(\rho) = \frac{1}{2} \sup \lset \<\rho, W'\> \bar W' \in \Spc \cap (-\Sp)^\circ \rset - \frac{1}{2}.
\end{align}
With the change of variables $W' = \mathds{1}-2W$, we obtain a common representation of the Lagrange duals of the robustnesses \cite{brandao_2005}:
\begin{align}\label{eq:gen_robustness_dual}
&\Rg\Sp(\rho) = \sup \lset -\<\rho, W\> \bar W \in \Spd \cap \DD^\circ \rset\\
\label{eq:standard_robustness_dual}
&\Rs\Sp(\rho) = \sup \lset -\<\rho, W\> \bar W \in \Spd \cap \Spc \rset.
\end{align}
It follows that strong Lagrange duality always holds for the robustness quantifiers by virtue of their being gauge functions. This will be considered in more generality in Sec. \ref{subsec:witness}.

We proceed by characterising the relation between the robustness measures and the previously introduced atomic gauges, as well as showing that the generalised robustness $\Rg\Sp$ always reduces to the vector atomic gauge $\A_\V$ on pure states.


\begin{boxed}{orange}
\begin{theorem}\label{prop:robustness_bound_as}
For any state $\rho \in \dom(\Rg\Sp)$, we have
\begin{equation}
  \A_\S(\rho) - 1 \geq \Rg\Sp(\rho).
\end{equation}
\end{theorem}
\end{boxed}
\begin{proof}
First, note that the generalised robustness can be expressed as
\begin{subequations}\label{eq:rob_dualH}
\begin{align}
  2 \Rg\Sp(\rho) + 1 &=  \A_{\Sp \cup (-\DD)}(\rho)\nonumber\\
  &= \sup \lset \<\rho, W\> \bar W \in \Spc,\; \<W, \sigma\>\geq - 1\;\forall \sigma\in\DD \rset\nonumber\\
  &= 2 \sup \lset \<\rho, W'\> \bar W' \in \Spc,\;W' \in \DD^*\rset - 1\label{eq:rob_line1}\\
  &= 2 \sup \lset \<\rho, W'\> \bar W' \in \Spc,\;W' \in \HH_{+}\rset - 1\label{eq:rob_line2}
\end{align}
\end{subequations}
where in \eqref{eq:rob_line1} we make the change of variables $W' = \frac12\left(W + \mathds{1}\right)$, and \eqref{eq:rob_line2} follows from the self-duality of the positive semidefinite cone $\HH_+$ and the fact that $\RR_+\DD = \HH_+$. We now have
\begin{equation}
\begin{aligned}
  \A_S (\rho) &= \sup \lset \left|\<\rho, Z\>\right| \bar Z \in \S^\circ \rset\\
  &\geq \sup \lset \left|\<\rho, Z\> \right|\bar Z \in \HH_+, Z \in \S^\circ \rset\\
  &= \sup \lset \left|\<\rho, Z\>\right| \bar Z \in \HH_+,\;\cbraket{a|Z|b}\leq 1\;\forall \ket{a},\ket{b}\in\V\rset\\
  &= \sup \lset \left|\<\rho, Z\>\right| \bar Z \in \HH_+,\;\braket{a|Z|a} \leq1\;\forall \ket{a}\in\V\rset\\
  &= \sup \lset \<\rho, Z\> \bar Z \in \HH_+, Z \in \Spc \rset\\
  &= \Rg\Sp(\rho) + 1\nonumber
\end{aligned}
\end{equation}
where the third equality follows from Lemma \ref{lemma:posdef}.
\end{proof}


\begin{boxed}{orange}
\begin{theorem}\label{prop:gen_robustness_pure_states}For any pure state $\proj\psi \in \dom(\Rg\Sp)$, the generalised robustness is equivalent to the vector atomic gauge for the set $\V$:
\begin{align}\Rg\Sp (\proj\psi) =  \A_\V (\ket\psi)^2 - 1
\end{align}
\end{theorem}
\end{boxed}
\begin{proof}
The fact that we have $\Rg\Sp (\proj\psi) \leq \A_\V(\ket\psi)^2 - 1$ follows from Thm. \ref{prop:robustness_bound_as}, since by Thm. \ref{prop:nucl_pure_states} we have $\A_\S(\proj\psi) = \A_\V(\ket\psi)^2$ for any $\ket\psi \in \sspan(\V)$.

To show that $\Rg\Sp(\proj\psi) \geq \A_V(\proj\psi)^2 - 1$, from the dual characterisation of the robustness we have
\begin{equation}
\begin{aligned}
  2 \Rg\Sp(\proj\psi) + 1 &= 2 \sup \lset \<\proj\psi, W'\> \bar W' \in \Spc,\;W' \in \HH_{+}\rset - 1\\
  &\geq 2 \sup \lset \<\proj\psi, \proj{w}\> \bar \proj{w} \in \Spc \rset-1\\
  &= 2 \sup \lset \cbraket{\psi|w}^2 \bar \<\proj{w}, \proj{v}\>\leq 1\;\forall \ket{v}\in\V\rset - 1\\
  &= 2 \sup \lset \cbraket{\psi|w}^2 \bar \ket{w} \in \V^\circ\rset - 1\\
  &= 2 \A_\V (\ket\psi) ^2 - 1.
\end{aligned}
\end{equation}
\end{proof}


Let us briefly discuss the implications of the above Theorem. It shows that the quantification of the robustness $\Rg\Sp$ for pure states, which can be a very nontrivial problem to solve for specific resource theories explicitly \cite{steiner_2003,clarisse_2006,cavalcanti_2005,ringbauer_2017},
 always reduces to an optimisation over the underlying Hilbert space only. This generalises results previously obtained in the resource theory of entanglement \cite{steiner_2003} and coherence \cite{piani_2016}, providing novel computable formulas and solving conjectures raised in the literature regarding the quantification of robustness measures of different resources \cite{clarisse_2006,johnston_2015,ringbauer_2017}, as we will discuss in detail in Sec. \ref{sec:app}.

 \begin{remark}
Notice that the proofs of Theorems \ref{prop:robustness_bound_as} and \ref{prop:gen_robustness_pure_states} do not rely on the fact that $\Sp$ consists of normalised density matrices. We can obtain a more general result in the following form.
\end{remark}
\begin{boxed}{white}
\begin{corollary}Let $\mathcal{Q} \coloneqq \conv \lset \proj{v} \bar \ket{v} \in \V \rset$ where $\V \subseteq \CC^d$ is a balanced and compact set. For any $\ket{x} \in \CC^d$, it then holds that
\begin{equation}\begin{aligned}
  \sup \lset \braket{x | W | x} \bar W \in \HH_+ \cap \mathcal{Q}^\circ \rset = \sup \lset \cbraket{x | w}^2 \bar \ket{w} \in \V^\circ \rset.
\end{aligned}\end{equation}
\end{corollary}
\end{boxed}

\begin{remark}
It is known that in the resource theory of entanglement, not only the generalised robustness, but also the standard robustness of entanglement reduces to the vector atomic gauge $\A_{\V}$ for pure states, i.e. $\Rs{\Sp}(\proj\psi) = \A_{\V}(\ket\psi)^2-1$ \cite{vidal_1999}. However, one can find that this does not hold for all quantum resources (see Sec. \ref{sec:magic} for an explicit example in the resource theory of magic states) --- it is then an interesting open question to characterise exactly when $\Rs{\Sp}$ is equivalent to $\A_{\V}$ on pure states. We remark that in any resource theory where $\Rs\Sp(\proj\psi) = \A_\V(\ket\psi)^2-1$ does hold, we have an extended hierarchy of quantifiers, all of which are equal on pure states:
\begin{equation}
  \A^\cup_\Sp(\rho) - 1\geq \Rs{\S_+} (\rho) \geq \A_{\S} (\rho) - 1\geq \Rg{\S_+} (\rho)
\end{equation}
which follows since $\A^\cup_\Sp$ is the largest convex function equal to $\A^2_\V$ on pure states.
\end{remark}


\begin{boxed}{orange}
\begin{proposition}\label{prop:robg_polar_equals}For any $P \in \HH_+$ it holds that
\begin{equation}
  \A^\circ_{\Sp\cup(-\DD)}(P) = \A^\circ_\Sp(P).
\end{equation}
\end{proposition}
\end{boxed}
\begin{proof}
By the positive semidefiniteness of $P$ we have
\begin{equation}\begin{aligned}
  \A^\circ_{\Sp\cup(-\DD)}(P) &= \sup \lset \<P, Z\> \bar Z \in \conv\left(\Sp \cup (-\DD)\right)\rset\\
  &= \max\left\{ 0,\; \sup \lset \<P, Z\> \bar Z \in \Sp \rset \right\}\\
  &= \sup \lset \<P, Z\> \bar Z \in \Sp \rset\\
  &= \A^\circ_\Sp(P).
\end{aligned}\end{equation}
\end{proof}
One can use the result of Prop. \ref{prop:robg_polar_equals} to give a bound on the value of $\Rg\Sp$ using the Cauchy-Schwarz inequality for $\A_{\Sp\cup(-\DD)}$, namely 
\begin{equation}\begin{aligned}
\Rg\Sp(\rho) \geq \frac{1}{2}\left(\frac{\Tr\left(\rho^2\right)}{\A^\circ_\Sp(\rho)} - 1\right).  
\end{aligned}\end{equation}
However, a tighter bound can be obtained.

\begin{boxed}{orange}
\begin{proposition}\label{prop:rob_dual_bounds}
For any state $\rho \in \dom(\Rg\Sp)$ it holds that
\begin{align}
  \Rg\Sp(\rho) \geq \frac{\Tr\left(\rho^2\right)}{\A^\circ_\Sp(\rho)} - 1.
\end{align}
\end{proposition}
\end{boxed}
\begin{proof}
First, notice that we do not need to assume $\A^\circ_\Sp(\rho) \neq 0$, since any positive semidefinite $\rho \in \dom(\Rg\Sp)$ satisfies $\supp(\rho) \subseteq \sspan(\V)$ and so we cannot have that $\<\rho, \sigma\> = 0 \;\forall \sigma \in \Sp$. From $\A^\circ_\Sp = \gamma_\Spc$ we then have that $\frac{\rho}{\A^\circ_\Sp(\rho)} \in \Spc$. Since $\rho \in \HH_+$, this is then a feasible solution to the dual formulation of the generalised robustness (Eq. \ref{eq:rob_dualH}) and so we have
\begin{equation}
  \Rg\Sp(\rho) \geq \< \rho, \frac{\rho}{\A^\circ_\Sp(\rho)} \> - 1
\end{equation}
as required.
\end{proof}



\section{Atomic gauges as resource quantifiers}\label{subsec:resource_quantifiers}

A full characterisation of a resource theory requires, in addition to the chosen set of free states without a given resource, the choice of a relevant set of \textit{free operations} which cannot generate a given resource \cite{brandao_2015}. For simplicity, in the following we will limit ourselves to the discussion of quantum operations acting as linear operators on the vector space of $d \times d$-dimensional Hermitian matrices $\HH$; the results can be generalised to maps between different vector spaces by suitably defining the set of free states in the input space as well as the set of free states in the output space (see e.g. \cite{gour_2017}).

The definition of what exactly constitutes the free operations is frequently dependent on the resource theory or the physical setting under consideration. The largest possible set of free operations are the \textit{resource non-generating operations} $\RNG$, such that for any $\Phi \in \RNG$ we have $\sigma \in \Spdd \,\Rightarrow\, \Phi(\sigma) \in \Spdd$. In particular, a trace-preserving resource non-generating operation acting on a free state always results in another free state. Often, a smaller subset of operations is considered: they are the \textit{stochastically resource non-generating operations} $\RF$, consisting of operations $\Lambda$ which admit a Kraus decomposition of the form $\Lambda(\rho) = \sum_n K_n \rho K^\dagger_n$ such that $\sigma \in \Spdd \,\Rightarrow\, K_n \sigma K^\dagger_n \in \Spdd \;\forall\,n$. The physical motivation for the choice of stochastically resource non-generating operations is that it guarantees that no resource can be created from a non-resource state in any possible measurement outcome. In the theory of theory of quantum coherence, for instance, they are called incoherent operations \cite{baumgratz_2014}. Even these operations are often considered to not reflect the physical restrictions sufficiently well, and much smaller subsets of physically relevant operations are used; for instance, in entanglement theory the set of operations of interest are the local operations and classical communication (\locc) \cite{vedral_1997,chitambar_2014}, and in coherence theory a common choice are the strictly incoherent operations \cite{winter_2016} or even smaller subsets \cite{chitambar_2016,marvian_2016,vicente_2017}.

Identifying $\conv(\Sp)$ with the set of free states and $\F$ with a chosen set of free maps (not necessarily channels), we define a valid measure for the given resource to be a function $M : \DD \to \mathbb{R}_+ \cup \{\infty\}$ which satisfies two basic criteria:

\begin{enumerate}[({C}1) ]
\item \textbf{Faithfulness}: $M(\rho) = 0$ if and only if $\rho \in \conv(\Sp)$.
\item \textbf{Monotonicity}: $M(\rho) \geq M(\Theta(\rho))$ for all completely positive trace-preserving (\cptp) operations $\Theta \in \F$.
\end{enumerate}
Often, additional conditions are imposed, the two most common ones being:
\begin{enumerate}[({C}1) ]
  \setcounter{enumi}{2}
  \item \textbf{Convexity}: $M\left(\sum_i t_i \rho_i\right) \leq \sum_i t_i M(\rho_i)$ for $t_i \in \RR_+$, $\rho_i \in \DD$, $\sum_i t_i = 1$.
  \item \textbf{Strong monotonicity}: $M(\rho) \geq \displaystyle \sum_n p_n M\left( \frac{K_n \rho K^\dagger_n}{p_n} \right)$
  where $\{K_n\}_n$ are Kraus operators corresponding to a quantum channel such that $K_n \cdot K^\dagger_n \in \F \;\forall\,n$, and the corresponding probabilities are given as $p_n = \Tr(K_n \rho K^\dagger_n)$.
\end{enumerate}

The choice of the free operations of interest $\F$ will, in general, not be unique --- however, we stress that (strong) monotonicity under a given class of operations implies (strong) monotonicity under any subset of this class, and so it suffices to investigate monotonicity under larger sets of operations. The notion of strong monotonicity was originally defined to reflect the monotonicity on average of the output states after a resource non-generating measurement (see e.g. \cite{vedral_1998,vidal_2000,baumgratz_2014}). Notice that, by definition of $\RNG$ and $\RF$, strong monotonicity under $\RNG$ coupled with convexity implies monotonicity under $\RF$.

\subsection{Properties of atomic gauges}

We now verify the criteria for the candidate measures: $\A_\S(\rho) - 1$, $\A^\cup_\Sp(\rho) - 1$, $\Rs\Sp(\rho)$, and $\Rg\Sp(\rho)$. 


\begin{boxed}{orange}
\begin{theorem}\label{prop:faithful}
A state $\rho \in \DD$ belongs to the set $\conv(\Sp)$ if and only if
\begin{equation}
  \A_\Sp(\rho) - 1 = \A_\S(\rho) - 1 = \A_\Spsym(\rho) -1 = \A^\cup_\Sp(\rho) - 1 = \Rs\Sp(\rho) = \Rg\Sp(\rho) = 0.
\end{equation}
That is, all of the considered measures are faithful.
\end{theorem}
\end{boxed}
\begin{proof}
It is helpful to recall that the gauge function of the set
\begin{equation}
  \mathcal{U} = \lset \ketbra{x}{y} \bar \braket{x|x} \leq 1,\, \braket{y|y} \leq 1 \rset
 \end{equation}
 is just the trace norm $\norm{\cdot}{1}$. Since $\Sp \subseteq \Spsym \subseteq \S \subseteq \mathcal{U}$, $\Sp \subseteq \Scup \subseteq \mathcal{U}$, and $\Sp \subseteq \Sp\cup(-\DD) \subseteq \mathcal{U}$ it follows that $\A_\mathcal{U}$ minorises all of the considered gauges, and so each of the gauges cannot be smaller than $1$ when $\rho \in \DD$ because $\norm{\rho}{1}=1$.

Then if $\rho \in \conv(\Sp)$, we have $\A_\Sp(\rho)=1$ by definition, and all the other gauges also have to be equal to 1 because they are majorised by $\A_\Sp$ (Eq. \eqref{eq:gauge_ineq}), which in turn means the robustnesses will be zero.

Conversely, assume that $\Rg\Sp(\rho) = 0$, which means that $\A_{\Sp\cup(-\DD)}(\rho) = 1$. This gives that $\rho \in \conv(\Sp\cup(-\DD))$, but since $\rho \in \DD$ by assumption and $\conv\left(\Sp\cup(-\DD)\right) \cap \DD = \conv(\Sp)$, we necessarily have $\rho \in \conv(\Sp)$. Noting by Thm. \ref{prop:robustness_bound_as} and the inequalities between the gauges that $\Rg\Sp(\rho)+1$ minorises all other gauges in consideration, it follows that any of the gauges being equal to $1$ implies that $\Rg\Sp$ is zero, proving the claim. 
\end{proof}


\begin{boxed}{orange}
\begin{proposition}All of the considered measures are convex.\end{proposition}
\end{boxed}
\begin{proof}Every gauge function of a convex set is convex.\end{proof}


To address the monotonicity of the robustness measures, we will consider a stronger notion of strong monotonicity under subchannels as has been employed e.g. in~\cite{piani_2016}, which readily implies the condition (C4) above. In particular, instead of a Kraus decomposition of a channel, we will consider a quantum instrument~\cite{davies_1970,ozawa_1984} which is a more general way of expressing probabilistic state transformations upon measurement: it is a collection of completely positive maps $\{\Lambda_i\}_i$ such that the overall transformation $\sum_i \Lambda_i$ is a valid quantum channel, and the input state $\rho$ transforms to the output state $\Lambda_i(\rho)$ with corresponding probability $p_i = \Tr(\Lambda_i(\rho))$.

\begin{boxed}{orange}
\setcounter{theorem}{19}
\begin{theorem}\label{thm:monotonic_any_gauge}
Let $\C$ be any compact set of states and let $\F \subseteq \RNG$ denote a chosen set of completely positive maps. Provided that the cone $\C\*\*$ is closed under the operations $\F$, i.e. $X \in \C\*\* \Rightarrow \Lambda(X) \in \C\*\*$ for all $\Lambda \in \F$, for any quantum channel $\Theta = \sum_i \Theta_i$ such that each $\Theta_i \in \F$ it holds that
\begin{equation}\begin{aligned}
  R_\Sp^\C (\rho) \geq \sum_i p_i \, R_\Sp^\C \!\left(\frac{\Theta_i}{p_i} \right)
\end{aligned}\end{equation}
for any $p_i \in \RR_+$ with $\sum_i p_i = 1$.
\end{theorem}
\end{boxed}
\begin{proof}
By definition of $\Gamma_{\Sp \cup (-\C)}$, write $\rho = \lambda^+ \sigma - \lambda^- \omega$ where $\sigma \in \Sp$, $\omega \in \C$ and $\lambda^+ + \lambda_- = \Gamma_{\Sp \cup (-\C)} (\rho)$. For each $i$, we then have
\begin{equation}\begin{aligned}
  \Theta_i (\rho) = \lambda^+ \Theta_i(\sigma) - \lambda^- \Theta_i (\omega).
\end{aligned}\end{equation}
But since $\Theta_i(\sigma) \in \Spdd$ and $\Theta_i(\omega) \in \C\*\*$ by assumption, this gives rise to a valid decomposition of $\Theta_i(\rho)$ as
\begin{equation}\begin{aligned}
  \Theta_i (\rho) = \lambda^+ \Tr \Theta_i(\sigma) \frac{\Theta_i(\sigma)}{\Tr \Theta_i(\sigma)} - \lambda^- \Tr \Theta_i(\omega) \frac{\Theta_i(\omega)}{\Tr \Theta_i(\omega)}. 
\end{aligned}\end{equation}
Hence,
\begin{equation}\begin{aligned}
  \Gamma_{\Sp \cup (-\C)} (\Theta_i(\rho)) \leq \lambda^+ \Tr \Theta_i(\sigma) + \lambda^- \Tr \Theta_i(\omega).
\end{aligned}\end{equation}
Using the positive homogeneity of $\Gamma_{\Sp \cup (-\C)}$ and the fact that $\sum_i \Theta_i$ is trace preserving, we obtain
\begin{equation}\begin{aligned}
  \sum_i p_i \Gamma_{\Sp \cup (-\C)} \! \left( \frac{\Theta_i(\rho)}{p_i} \right) &= \sum_i \Gamma_{\Sp \cup (-\C)} (\Theta_i(\rho))\\
  &\leq \sum_i \left[ \lambda^+ \Tr \Theta_i(\sigma) + \lambda^- \Tr \Theta_i(\omega) \right]\\
  &= \lambda^+ + \lambda^-\\
  &= \Gamma_{\Sp \cup (-\C)}(\rho).
\end{aligned}\end{equation}
Recalling that $R_\Sp^\C$ is a linear function of $\Gamma_{\Sp \cup (-\C)}$ gives the same monotonicity relation for the robustness.
\end{proof}

\begin{boxed}{orange}
\setcounter{theorem}{17}
\begin{corollary}The robustness measures $\Rg\Sp$ and $\Rs\Sp$ satisfy both monotonicity and strong monotonicity under $\RNG$: the former can be seen by choosing a single channel $\Theta_i = \Theta$ in the above, and the latter by choosing $\Theta_i = K_i \cdot K^\dagger_i$.
\end{corollary}
\end{boxed}

\begin{remark}
The Theorem follows analogously when $\C$ is not a set of states, but any set of Hermitian operators of fixed trace.
\end{remark}

\begin{boxed}{orange}
\begin{theorem}\label{thm:monotone_conv}The atomic gauges $\A_\S$ and $\A^\cup_\Sp$ satisfy strong monotonicity under $\RNG$, and therefore monotonicity under stochatically resource non-generating maps $\RF$.\end{theorem}
\end{boxed}
\begin{proof}
Consider $\A_\S$ first, and take a channel with a Kraus decomposition as $\Lambda(\rho) = \sum_n K_n \rho K^\dagger_n$ such that $\sigma \in \Spdd \,\Rightarrow\, K_n \sigma K^\dagger_n \in \Spdd \;\forall\,n$. By definition of the gauge function, any state $\rho$ can be written as $\rho = \sum_{i} c_{i} \ketbra{v_i}{u_i}$ for some coefficients $c_{i} \in \RR_+$ with $\sum_i c_i = \Gamma_\S(\rho)$ and states $\ket{v_i} \in \V$, $\ket{u_i} \in \V$.

Note first that for all pure states $\ket{v_i}, \ket{u_i} \in \V$ we have $K_n \ket{v_i} = q_{ni} \ket{v_{ni}}$ and $K_n \ket{u_i} = s_{ni} \ket{u_{ni}}$ for some $\ket{v_{ni}}, \ket{u_{ni}} \in \V$ and $q_{ni},  s_{ni} \in \RR_+$ as the complex phase can be absorbed into the states $\ket{v_{ni}}, \ket{u_{ni}}$. For each $i$, we then have
\begin{equation}\begin{aligned}\label{eq:Kraus_mon}
  \sum_n \Gamma_S \left( K_n \ketbra{v_i}{u_i} K^\dagger_n\right) &= \sum_n \Gamma_S \left( q_{ni} s_{ni} \ketbra{v_{ni}}{u_{ni}} \right)\\
  &= \sum_n q_{ni} s_{ni} \, \Gamma_S \left( \ketbra{v_{ni}}{u_{ni}} \right)\\
  &= \sum_n q_{ni} s_{ni}\\
  &\leq \sqrt{ \sum_n q_{ni}^2 }\sqrt{ \sum_n s_{ni}^2 }\\
  &= 1
\end{aligned}\end{equation}
where the inequality follows by the Cauchy-Schwarz inequality, and the last equality is a consequence of the fact that $\sum_n K_n \cdot K^\dagger_n$ constitutes a valid quantum channel and thus
\begin{equation}\begin{aligned}
  1 = \Tr\left(\sum_n K_n \proj{v_i} K^\dagger_n\right) = \Tr\left( \sum_n p^2_{ni} \proj{v_{ni}} \right) = \sum_n p^2_{ni}
\end{aligned}\end{equation}
and similarly for $\ket{u_i}$.

For the state $\rho$, this then gives
\begin{equation}\begin{aligned}
\sum_n p_n \, \Gamma_\S \left( \frac{K_n \rho K^\dagger_n}{p_n} \right) &= \sum_n \Gamma_\S \left( K_n \rho K^\dagger_n\right) \\
&=  \sum_n \Gamma_\S \left( \sum_i c_i K_n \ketbra{v_i}{u_i} K^\dagger_n \right)\\
  &\leq \sum_{n} \sum_i c_i  \, \Gamma_\S \left( K_n \ketbra{v_i}{u_i} K^\dagger_n \right)\\
  &\leq \sum_i c_i\\
  &= \Gamma_\S(\rho),
\end{aligned}\end{equation}
where in the first line we used the positive homogeneity of $\Gamma_\S$, in the third line its subadditivity and homogeneity, and in the fourth line we used Eq.~\eqref{eq:Kraus_mon}.

The case of $\A^\cup_\Sp$ can be shown analogously, or even more straightforwardly: writing $\rho = \sum_i r_i \proj{\psi_i}$ in the pure-state decomposition such that $\A^\cup_\Sp \!(\rho)= \sum_i r_i \Gamma^2_\V (\ket{\psi_i})$, we have
\begin{equation}\begin{aligned}
  \sum_n p_n \, \A^\cup_\Sp \!\left( \frac{K_n \rho K^\dagger_n}{p_n} \right) &= \sum_n \A^\cup_\Sp \! \left( K_n \rho K^\dagger_n\right) \\
&=  \sum_n \A^\cup_\Sp \left( \sum_i r_i K_n \proj{\psi_i} K^\dagger_n \right)\\
  &\leq \sum_{n} \sum_i r_i  \, \A^\cup_\Sp \!\left( K_n \proj{\psi_i} K^\dagger_n \right)\\
  &\leq  \sum_i r_i \, \A^\cup_\Sp\!(\proj{\psi_i})\\
  &= \A^\cup_\Sp (\rho)
\end{aligned}\end{equation}
where the last inequality follows from the strong monotonicity of $\A^\cup_\Sp$ for pure states, i.e.
\begin{equation}\begin{aligned}
  \sum_n \A^\cup_\Sp \!\left( K_n \proj{\psi_i} K^\dagger_n \right) \leq \A^\cup_\Sp (\proj{\psi_i}),
\end{aligned}\end{equation}
 which itself is a direct consequence of the strong monotonicity of $\Rg\Sp$ (Thm.~\ref{thm:monotonic_any_gauge}) and the fact that $\A^\cup_\Sp (\proj{\psi_i}) = \Rg\Sp(\proj{\psi_i}) + 1$ for any pure state (Thm.~\ref{prop:gen_robustness_pure_states}).
\end{proof}
\setcounter{theorem}{20}

We have thus verified that the robustness measures form a valid class of strong monotones in any resource theory, since they satisfy monotonicity under the largest class of free operations $\RNG$. The measures $\Gamma_\S$ and $\Gamma^\cup_\Sp$, on the other hand, constitute valid monotones as long as $\F \subseteq \RF$ (and indeed they are, in general, not monotonic under $\RNG$~\cite{bu_2017-1}).

\begin{remark}Instead of choosing the resource quantifier as $M = \A_\C - 1$ for one of the gauge functions $\A_\C$, one can instead consider $M = f \circ \A_\C$ for any monotonically non-decreasing function $f$ on the interval $[1,\infty)$ s.t. $f(1)=0$. Any choice of a convex $f$ then preserves the convexity of the measure, and any choice of a concave $f$ preserves the strong monotonicity --- a common choice of a concave function is $f = \log$ \cite{vidal_2002,plenio_2005,rana_2017}.
\end{remark}


\subsection{Dual characterisation and resource witnesses}\label{subsec:witness}

In quantum resource theories, elements of the dual cone $\Spd$ are often called \textbf{witnesses} of the given resource. The notion of witnesses is a fundamental concept that found a variety of uses in the characterisation, detection, and quantification of quantum resources \cite{horodecki_1996-1,terhal_2002,brandao_2005,guhne_2009}.

The set of witnesses can also be described in terms of the polar set $\Spc$ --- the equivalence is made explicit by noting that a witness can be obtained from any $S \in \Spc$ simply by considering $\mathds{1}-S \in \Spd$. This gives an intuitive interpretation of the polar gauge in this sense:
\begin{boxed}{orange}
\begin{proposition}For any $X \in \HH$ it holds that
\begin{equation}
  \A^\circ_\Sp(X) = \inf \lset \lambda \in \RR_+ \bar \lambda \mathds{1} - X \in \Spd \rset.
\end{equation}
\end{proposition}
\end{boxed}
\begin{proof}
Recalling that $\A^\circ_\Sp = \A_\Spc = \gamma_\Spc$, we have $\A^\circ_\Sp(X) = \inf \lset \lambda \in \RR_+ \bar X \in \lambda \Spc \rset$ and the result follows since $\<\mathds{1}, \sigma\> = 1 \;\forall \sigma \in \Sp$.
\end{proof}
\begin{remark} Notice that if $X$ is such that $\< \sigma, X \> \leq 0 \;\forall \sigma \in \Sp$, then $\A^\circ_\Sp(X) = 0$. In particular, a positive semidefinite matrix can have $\A^\circ_\Sp(X) = 0$ if and only if $X \in \Sperp$, where $\Sperp = \Spd \cap (-\Sp)\*$ is the orthogonal complement of $\Sp$. Note that $\Sperp \cap \HH_+ = \{0\}$ when $\sspan(\V) = \CC^d$.
\end{remark}

The crucial property of witnesses applied to quantum resources is that, since $\conv(\Sp)$ is a convex and closed set, by the strongly separating hyperplane theorem \cite[11.4]{rockafellar_1970} for every $\rho \notin \conv(\Sp)$ there exists a witness $W \in \Spd$ such that $\< \rho, W\> < 0$, thus \textit{detecting} a given resource. On the other hand, if $\rho \in \conv(\Sp)$, then we necessarily have $\< \rho, W \> \geq 0 \;\forall W \in \Spd$. This leads to a natural formalism of witness-based measures, quantifying how much a given state can violate the condition $\< \rho, W \> \geq 0$: one constructs a general witness-based measure as \cite{brandao_2005}
\begin{equation}\label{eq:witness_measure}
  \VV(\rho) = \sup \lset -\< \rho, W \> \bar W \in \Spd \cap \C \rset 
\end{equation}
where $\C$ is a set representing some additional constraints on the set of witnesses under consideration. This approach is particularly useful in experimental settings, allowing for the detection and quantification of resources without the need for full state tomography \cite{guhne_2007-1,eisert_2007}.

Many of the gauge functions considered in this work can be written in this form with a suitable choice of $\C$ (up to a constant), which can be seen from their dual representation. For example, the standard robustness corresponds to $\VV$ with the choice $\C = \Spc$, and the generalised robustness to $\C = \DD^\circ = \lset X \bar X \cleq \mathds{1} \rset$. More generally, we can establish the following equivalence.
\begin{boxed}{orange}
\begin{proposition}\label{prop:witness_gauge}
Recall that $\HH_1 = \lset X \in \HH \bar \<\mathds{1}, X\> = 1 \rset$. If $\C$ can be expressed as $\C = \X^\circ$ for some subset $\X \subseteq \HH_1$, then
\begin{equation}
  \VV(\rho) = \frac{1}{2}\left(\A_{\Sp\cup(-\X)}(\rho) - 1\right)
\end{equation}
for any $\rho \in \DD$. Likewise, if $\C = \Y^\circ$ for some $\Y \subseteq -\HH_1$, then
\begin{equation}
  \VV(\rho) = \A_{\Sp\cup\left(-\frac{1}{2}\Y\right)}(\rho) - 1.
\end{equation}
\end{proposition}
\end{boxed}
\begin{proof}
Take the standard expression for the witness-based measure in Eq. \eqref{eq:witness_measure} and apply the change of variables $W' = \mathds{1} - 2W$ in the case of  $\X$, or $W' = \mathds{1} - W$ in the case of $\Y$. The result then follows from the dual characterisation of the gauge functions.
\end{proof}
One can note a similarity of the above forms to the generalised base gauges considered in section \ref{sec:robustness}.

In the cases considered in Prop. \ref{prop:witness_gauge}, strong Lagrange duality always holds by virtue of the dual representation of atomic gauges. For more general sets $\C$ one can note that since $\Sp$ is bounded, we have $0 \in \interior(\Spc)$ \cite[14.5.1]{rockafellar_1970}, and it follows that $\mathds{1} \in \interior(\Spd)$. This fact is useful in showing that strong Lagrange duality holds for a given measure --- by Slater's condition \cite{boyd_2004}, strong duality holds if there exists a witness $W \in \relint(\Spd \cap \C)$. It is then sufficient to show the existence of a witness in $\relint(\C)$ which is in the neighbourhood of $\mathds{1}$.

An example of a witness-based measure is obtained by taking $\VV$ with $\C = \lset X \bar \< \mathds{1}, X \> \leq d \rset$ \cite{brandao_2006}. Noting that this corresponds to $\C = \X^\circ$ with $\X = \left\{ \frac{\mathds{1}}{d} \right\}$, we obtain
\begin{equation}\VV(\rho) = R^{\{\mathds{1}/{d}\}}_\Sp(\rho) = \inf \lset s \in \RR_+ \bar \rho + s \frac{\mathds{1}}{d} \in \Spdd \rset\end{equation}
which defines the \textit{random robustness} \cite{vidal_1999}. Random robustness has full domain iff $\frac{\mathds{1}}{d} \in \relint_\DD(\Sp)$, and it corresponds to the gauge function $\frac{1}{2}\left(\A_{\Sp\cup\left\{-\mathds{1}/d\right\}} - 1 \right)$. Alternatively, one can think of it as the atomic gauge function of $\Sp$ under a reparametrisation such that $\frac{\mathds{1}}{d}$ is the origin.  Note that the random robustness is not a monotone under the general free operations $\RNG$, but by Theorem \ref{thm:monotonic_any_gauge} it can be noted to be faithful and strongly monotonic under unital resource non-generating operations (i.e. $\Phi \in \RNG$ such that $\Phi(\mathds{1}) = \mathds{1}$). By the set inclusion of the corresponding unit balls we get $R^{\{\mathds{1}/{d}\}}_\Sp(\rho) \geq \Rs\Sp(\rho)$.

Another common choice is to consider $\VV$ with $\C = (-\DD)^\circ = \lset X \bar X \cgeq -\mathds{1} \rset$, which by strong Lagrange duality is equal to the so-called \textit{best free approximation} $\operatorname{BFA}$, generalising the best separable approximation \cite{lewenstein_1998}:
\begin{equation}\label{eq:bfa}
\operatorname{BFA}_\Sp(\rho) = \min_{\sigma \in \conv\Sp} \lset 1-\lambda \bar \rho \cgeq \lambda \sigma \rset.
\end{equation}
Here, the optimal $\lambda$ is the largest weight that a free state can take in a convex decomposition of $\rho = \lambda \sigma + (1-\lambda) \omega$ where $\sigma \in \conv(\Sp)$ and $\omega \in \DD$. This expression is then equivalent to the gauge function $\A_{\Sp\cup\left(\frac{1}{2}\DD\right)} - 1$. Since the cone generated by the set $\Sp\cup\left(\frac{1}{2}\DD\right)$ is closed under resource non-generating operations, we have that the best free approximation is a faithful strong monotone under $\RNG$ by Theorem \ref{thm:monotonic_any_gauge}. Note, however, that it has some unusual properties for a resource monotone; for example, any resourceful pure state $\ket\psi \notin \V$ has $\operatorname{BFA}_\Sp(\proj\psi) = 1$, which fails to account for the fact that some pure states can be strictly more resourceful than others.

\subsection{Relations with distance-based measures}

One can define a faithful quantifier of any resource simply by considering the distance to the set of free states. That is, we can define
\begin{equation}
  D_\Sp(\rho) = \enskip\inf_{\mathclap{\sigma \in \conv(\Sp)}}\enskip D(\rho, \sigma)
\end{equation}
for some quasi-metric $D$ contractive under completely positive trace-preserving maps \cite{vedral_1997,bengtsson_2007}. Any such measure is then a (weak) monotone under resource non-generating operations. Although the representation is appealing from an intuitive point of view, the measures defined in this way are frequently difficult to evaluate and investigate in practice, and few results about their properties such as strong monotonicity are known.

The distance-based quantifiers do not fit into the gauge function formalism directly, but we can nevertheless obtain relations between the gauge-based and distance-based measures in several cases --- we will consider some representative examples of such quantifiers.

\subsubsection{Trace distance and other gauge-based distances}

A commonly encountered case is when $D$ is itself based on a gauge function --- this includes, for instance, the fundamental measure of trace distance obtained for $D(\rho,\sigma) = \norm{\rho-\sigma}{1}$. In general, one can consider
\begin{equation}
  D_\Sp(\rho) = \enskip\inf_{\mathclap{\sigma \in \conv \Sp}}\quad \A_\C(\rho-\sigma)
\end{equation}
where we will take $\C\subseteq \HH$ for simplicity, but the same considerations apply to more general sets of complex matrices with the inner product $\Re\<\cdot,\cdot\>$. 
To see the difference between the distance-based measures and gauge functions, we can consider the dual form of the general quantifier $D_\Sp$:
\begin{equation}\begin{aligned}
  D_\Sp(\rho) &= \inf_{\sigma \in \conv(\Sp)} \sup_{W \in \C^\circ} \< W, \rho - \sigma \>\\
  &= \sup \lset \<W, \rho \> - \sup_{\sigma \in \Sp} \<W, \sigma\> \bar W \in \C^\circ \rset
\end{aligned}\end{equation}
which follows by Sion's minimax theorem (or can be equivalently derived by considering the Lagrange dual of the original problem explicitly). We emphasise that the function $\mu_\Sp(W) \coloneqq \sup_{\sigma \in \Sp} \<W, \sigma\>$ is, in general, different from the polar gauge $\A^\circ_\Sp(W) = \left[\mu_\Sp(W)\right]_+$. This representation can nevertheless be helpful since the polar gauges, and as a result the function $\mu_\Sp$, can be relatively easy to characterise (see e.g. Sec. \ref{sec:app}).

It is explicit from the above representation that, although distance measures can be based on gauge functions, they are not gauges themselves. An easy way to modify any gauge-based distance measure $D_\Sp$ to turn it into a gauge function is simply to consider instead the distance to the cone generated by $\Sp$:
\begin{equation}\begin{aligned}
   D'_\Sp(\rho) \coloneqq& \enskip\inf_{\mathclap{X \in \Spdd}}\; \A_\C(\rho-X)\\
  =& \sup \lset \<W, \rho \> - \sup_{X \in \Spdd} \<W, X\> \bar W \in \C^\circ \rset\\
  =& \sup \lset \<W, \rho \> \bar W \in \C^\circ \cap (-\Spd) \rset\\
  =& \Gamma_{\C \cup \Spdd} (\rho).
\end{aligned}\end{equation}

As an explicit example, consider first the trace distance $T_\Sp(\rho)$, which corresponds to $D_\Sp$ with the choice $\C = \lset X \bar \norm{X}{1}\leq 1 \rset$ and thus gives
\begin{equation}
  T_\Sp(\rho) = \sup \lset \<W, \rho\> - \mu_\Sp (W) \bar \norm{W}{\infty} \leq 1 \rset.
\end{equation}
An alternative formulation of the trace distance can be obtained by recalling that the trace norm $\norm{\cdot}{1}$ admits a representation as the base norm $\A_{\DD\cup(-\DD)}$ in the space of Hermitian matrices, allowing us to write
\begin{equation}\begin{aligned}
    T_\Sp(\rho) &= \inf_{\sigma \in \conv(\Sp)} \sup \lset \<W, \rho - \sigma\>  \bar W \in \DD^\circ \cap (-\DD)^\circ \rset\\
    &= \inf_{\sigma \in \conv(\Sp)} \sup \lset \<W', \rho - \sigma\> - \<\mathbbm{1}, \rho\> + \<\mathbbm{1}, \sigma\> \bar W' \in \left(\frac{1}{2}\DD\right)^\circ \cap \DD\* \rset\\
    &= 2 \sup \lset \<W', \rho\> - \A^\circ_\Sp(W') \bar \mathbbm{1} \cgeq W' \cgeq 0 \rset
\end{aligned}\end{equation}
where in the second line we make the change of variables $W' = W + \mathbbm{1}$, and the last equality follows because $\mu_\Sp(W') = \A^\circ_\Sp(W')$ for any $W'\in\DD\* = \HH_+$. 
The trace distance can then be turned into a gauge function by considering $T'_\Sp(\rho) = \min_{X \in \Spdd} \norm{\rho-X}{1}$, a quantity known as the \textit{modified trace distance}. We can relate it to the generalised robustness as
\begin{equation}\begin{aligned}
  T'_\Sp(\rho) &= \sup \lset - \<\rho, W\> \bar W \in \DD^\circ \cap (-\DD^\circ) \cap \Spd \rset\\
  &\leq \sup \lset -\<\rho, W\> \bar W \in \DD^\circ \cap \Spd \rset\\
  &= \Rg\Sp(\rho).
\end{aligned}\end{equation}
The advantage of using $T'_\Sp$ over $T_\Sp$ is, once again, that the modified trace distance is a gauge function, and therefore immediately inherits all of the desirable properties of gauges in resource quantification that we discussed in Sec. \ref{subsec:resource_quantifiers}. As an example, a long-standing conjecture that the trace distance of entanglement is a strong monotone \cite{eisert_2003} was recently put to rest by employing results from the resource theory of coherence to show that the trace distance does \textit{not} satisfy strong monotonicity under the relevant stochastically resource non-generating operations \cite{yu_2016} --- this gives an interesting example of a case where it becomes necessary to consider distances with respect to the unnormalised cone $\Spdd$ in order to ensure strong monotonicity.

\subsubsection{Geometric measures}

The family of \textit{geometric measures} with respect to $\S_+$ is defined by noting that, for pure states, the quantity $\A^\circ_\V(\ket\psi)$ effectively quantifies the resource contained in a state by measuring the largest possible overlap with a free state \cite{shimony_1995,wei_2003}, reaching its maximal value $1$ only on the set of free states. One can then define a resource quantifier simply as
\begin{equation}\begin{aligned}
  G_{\S_+}(\proj\psi) &\coloneqq 1 - \A^{\circ}_\V(\ket\psi)^2.
\end{aligned}\end{equation}
This function can now be extended to all mixed states through the convex roof, giving
\begin{equation}\begin{aligned}
  G_{\S_+}(\rho) &\coloneqq \left(1 - \A^{\circ 2}_\V\right)^\cup (\rho)
  = 1 - \A^{\circ \cap}_\Sp(\rho)\\
  &= 1 - \max \lset \sum_i p_i\, \A^\circ_\V\left(\ket{\psi_i}\right)^2 \bar \rho = \sum_i p_i \proj{\psi_i},\;p_i \in \RR_+,\;\sum_i p_i = 1 \rset.
\end{aligned}\end{equation}
Since $G_{\S_+}(\proj\psi) = 1 - \max_{\ket{\eta} \in \V} \cbraket{\psi|\eta}^2 = 1 - \sup_{\sigma \in \conv\Sp} F(\proj\psi, \sigma)$, one could expect this measure to be related to the fidelity-based distance quantifier:
\begin{equation}
  G'_{\S_+}(\rho) \coloneqq 1 - \;\, \max_{\mathclap{\sigma \in \conv(\Sp)}} \;\;\, F(\rho,\sigma).
\end{equation}
Surprisingly, the two quantities are in fact equal, which can be seen as follows.
\begin{boxed}{white}
\begin{proposition}[Streltsov et al. \cite{streltsov_2010}] The geometric measure $G_\Sp$ and the fidelity-based distance measure $G'_\Sp$ are equal. That is,
\begin{equation}
  \A^{\circ \cap}_\Sp(\rho) = \enskip\sup_{\mathclap{\sigma \in \conv(\Sp)}} \enskip\norm{\sqrt{\vphantom{\sigma}\rho}\sqrt{\vphantom{\rho}\sigma}\,}{1}^2.
\end{equation}
\end{proposition}
\end{boxed}
The proof, applied to any set $\Sp$ in the present formalism, follows straightforwardly from \cite{streltsov_2010}. The result is remarkable in that it reduces the quantification of the convex roof--based measure $G_\Sp$ to a potentially much less complicated convex optimisation problem.

$\A^{\circ \cap}_\Sp$ is concave and therefore is larger on $\DD$ than any convex function equal to it on the extremal points of the set, including in particular the gauge $\A^\circ_\Sp$ itself. Noting the equivalence between dual gauges (Props. \ref{prop:gauge_polars_equals} and \ref{prop:robg_polar_equals}), we thus obtain the bound
\begin{equation}
  G_\Sp(\rho) \leq 1 - \A^{\circ}_\Sp(\rho) = 1 - \A^{\circ}_\Spsym(\rho) = 1 - \A^\circ_\S(\rho) = 1 - \A^\circ_{\Sp\cup(-\DD)}(\rho)
\end{equation}
which holds for any density matrix $\rho$. From Prop. \ref{prop:rob_dual_bounds} we then have that
\begin{equation}
  \Rg\Sp(\rho) \geq \frac{\Tr(\rho^2)}{\A^\circ_\Sp(\rho)} - 1 \geq \frac{\Tr(\rho^2)}{1 - G_\Sp(\rho)} - 1.
\end{equation}
for any $\rho \in \dom(\Rg\Sp)$. This in particular establishes the relation
\begin{equation}\label{eq:robustness_geometric}
  \Rg\Sp(\proj\psi) \geq \frac{1}{1-G_\Sp (\proj\psi)} - 1 = \frac{G_\Sp (\proj\psi)}{1 - G_\Sp (\proj\psi)}
\end{equation} 
for all $\ket\psi \in \sspan(\V)$, which generalises a known property of the robustness of entanglement \cite{cavalcanti_2006}.


\subsubsection{Relative entropy}

Let $D(\rho\|\sigma)$ denote the quantum relative entropy, defined as
\begin{equation}
  D(\rho\|\sigma) = \begin{cases} \Tr \left(\rho \log \rho - \rho \log \sigma\right) & \text{if } \supp(\rho) \subseteq \supp(\sigma),\\
  \infty & \text{otherwise}.\end{cases}
 \end{equation}
This quantity is neither symmetric nor does it satisfy the triangle inequality, but it nevertheless can be treated as a quasi-metric and used as a measure to faithfully quantify a given quantum resource \cite{vedral_1997,bengtsson_2007}. A quantifier of interest is then the \textit{relative entropy with respect to $\Sp$},
\begin{equation}
  S_\Sp (\rho) = \enskip\inf_{\mathclap{\sigma \in \conv\Sp}}\enskip D(\rho\|\sigma).
\end{equation}
Note that $\dom(S_\Sp) = \dom(\Rg\Sp)$.

One can obtain useful relations between this quantity and atomic gauges, generalising results for entanglement and coherence theories found for example in  \cite{datta_2009, zhu_2017, rana_2017}. The connection between the two frameworks is made through quantities called the max- and min-relative entropy, defined as \cite{datta_2009,dupuis_2012}
\begin{equation}\begin{aligned}
  D_{\max} (\rho \| \sigma) &= \inf \lset s \in \RR_+ \bar \rho \cleq 2^s \sigma \rset = \log \inf \lset s \in \RR_+ \bar \rho \cleq s \sigma \rset\\
  D_{\min} (\rho \| \sigma) &= - 2 \log F(\rho,\sigma)
\end{aligned}\end{equation}
From the definition of the generalised robustness $\Rg\Sp$, it then follows that
\begin{equation}
  \enskip\inf_{\mathclap{\sigma \in \conv(\Sp)}}\quad D_{\max}(\rho\|\sigma) = \log \left(1 + \Rg\Sp(\rho)\right),
\end{equation}
making the $D_{\max}$-based measure strongly monotonic under resource non-ge\-ne\-ra\-ting operations. Using the relation $D_{\min} (\rho \| \sigma) \leq D (\rho \| \sigma) \leq D_{\max} (\rho \| \sigma)$ \cite{datta_2009,dupuis_2012}, we then have
\begin{equation}
  G_\Sp(\rho) \leq -2 \log(1-G_\Sp(\rho)) \leq S_\Sp (\rho) \leq \log\left(1 + \Rg\Sp(\rho)\right) \leq \log \A_\S (\rho)
\end{equation}
where the first inequality comes from the fact that $-2\log(x) \geq 1-x \; \forall x\in[0,1]$, and the last inequality from Theorem \ref{prop:robustness_bound_as}. This establishes a quantitative relation between some non-convex resource monotones and the relative entropy. In the resource theory of quantum coherence, it has been conjectured that the inequality $S_\Sp(\rho) \leq \A_\S(\rho) - 1$ also holds \cite{rana_2016,rana_2017}. Following \cite{rana_2017}, we can use the fact that $x \geq \log (1+x) \,\forall x \geq 1$ to similarly establish
\begin{align}
  &S_\Sp(\rho) \leq \Rg\Sp(\rho) \leq \A_\S(\rho) - 1 &&\forall\, \rho \text{ s.t. } \Rg\Sp(\rho) \geq 1\\
  &S_\Sp(\rho) \leq \Rg\Sp(\rho)\log e \leq \left(\A_\S(\rho) - 1\right)\log e \, &&\forall\, \rho,
\end{align}
and it is an interesting open question whether one can tighten the relation.


\section{Applications}\label{sec:app}

We now show how to apply the formalism introduced in the previous sections to two representative quantum resources --- entanglement and coherence --- as well as the recently established resource theory of magic states, demonstrating the universality of the results obtained in this work. We introduce new gauge-based quantifiers as well as show that many known measures of the three resources fit into the gauge function framework, allowing for their characterisation in this formalism. An important point to note here is that the properties of the measures such as strong monotonicity under relevant free operations, faithfulness, as well as quantitative bounds and relations, all follow straightforwardly from our discussion in Sec. \ref{sec:quantum} and \ref{subsec:resource_quantifiers}. Since these properties are often not easy to show explicitly for a given measure, one can benefit from exploiting the fact that these quantifiers are in fact atomic gauges. 

We begin with the discussion of quantum coherence, as the introduced measures will form a basis for our investigation of bipartite entanglement.

\subsection[k-coherence]{\texorpdfstring{$k$}{k}-coherence}

The resource-theoretic framework of quantum coherence in finite-dimensional systems was established relatively recently \cite{aberg_2006,gour_2008,levi_2014,baumgratz_2014} (see \cite{streltsov_2017} for a comprehensive overview). An extension of this concept to a hierarchy of $k$ levels of coherence has been considered in \cite{killoran_2016,chin_2017,chin_2017-1,theurer_2017,regula_2018-2} and was formalised as the resource theory of $k$-coherence in \cite{ringbauer_2017}. We remark that many of the methods introduced in this section can be used with non-orthogonal bases, thus applying also to resource theories of superposition which generalise quantum coherence \cite{killoran_2016,regula_2018-2,theurer_2017,mukhopadhyay_2017}.

Let us begin with the definition of the most general notion of quantum coherence. Since quantum coherence is based on superposition, it is fundamentally a basis-dependent concept. The set of free pure states in this resource theory, which we will denote as $\I$, corresponds to state vectors which have only one coefficient in a fixed orthonormal basis $\{\ket{i}\}$, that is, each free pure state vector is a normalised scalar multiple of a basis vector. The free density matrices, defined in the usual way as the convex hull of $\lset \proj\psi \bar \ket\psi \in \I \rset$, are called the \textit{incoherent} states. Any state which cannot be written as a convex combination of incoherent pure states is then \textit{coherent}, and can be used as a resource.

In some applications, however, not every coherent state is useful for the considered physical tasks \cite{killoran_2016,chin_2017,chin_2017-1,theurer_2017,regula_2018-2,ringbauer_2017} --- in these circumstances, a fine-grained quantification of quantum coherence is necessary. Given a fixed orthonormal basis $\{\ket{i}\}$, the \textit{coherence rank} $\CR(\ket\psi)$ is defined as the cardinality of the state vector $\ket\psi$, that is, the number of non-zero coefficients in this basis. The set of free pure states is then given by $\I^k = \lset \ket{\phi} \bar \CR(\ket\psi) \leq k,\; \braket{\psi|\psi} = 1 \rset$ and we extend this to mixed states as done previously; for the simplicity of notation, we will hereafter work with the convex hull directly:
\begin{equation}
\Ckp = \conv\lset \proj\psi \bar \ket\psi \in \I^k \rset.
\end{equation}
Any state $\rho \in \Ckp$ is then a free state in the resource theory of $k+1$-coherence, with all states $\rho \notin \Ckp$ being $k+1$-coherent. Note that $k=1$, that is the quantification of $2$-coherence, corresponds to the standard notion of quantum coherence, where the only resource-free states are convex combinations of the basis projectors; also, note that $\C^d_+ = \DD$. Let us also define the corresponding set $\Ck =  \conv\lset \ket\psi\bra\phi \bar \ket\psi \in \I^k,\, \ket\phi\in\I^k \rset$ as before.

\subsubsection{\texorpdfstring{$k=1$ and $k=d$}{k=1 and k=d}}

The resource theory of quantum coherence ($k=1$) has been investigated in many works, and a variety of quantifiers have been considered \cite{baumgratz_2014,streltsov_2015,yuan_2015,winter_2016,napoli_2016,piani_2016,qi_2017}. Applying the formalism of this work to the states $\Ckp$ with $k=1$, one can notice that we have:
\begin{equation}\begin{aligned}
  \A_{\I^1}(\ket\psi) &= \inf \lset \sum_i c_i \bar \ket\psi = \sum_i c_i \ket{v_i},\, c_i \in \RR_+,\, \ket{v_i} \in \I^1 \rset\\
  &= \inf \lset \sum_i |c_i| \bar \ket\psi = \sum_i c_i \ket{i},c_i \in \CC \rset\\
  &= \lnorm{\ket\psi}{1}.
\end{aligned}\end{equation}
Following in a similar way for the other quantifiers, we obtain several known norms and gauges for Hermitian matrices, many of which have been used as coherence measures:\vspace{10pt}
\begin{adjustbox}{center}
\begin{tabular}{@{} *2l  *2l @{}}  \toprule
Gauge & Also known as & Gauge & Also known as\\\midrule
$\A_{\I^1}$ & vector $\ell_1$ norm $\lnorm{\cdot}{1}$ & $\A_{\I^d}$ & vector $\ell_2$ norm $\lnorm{\cdot}{2}$\\
$\A^\circ_{\I^1}$ & vector infinity norm $\lnorm{\cdot}{\infty}$ & $\A^\circ_{\I^d}$ & vector $\ell_2$ norm $\lnorm{\cdot}{2}$\\
$\A_{\C^1}$ & element-wise $\ell_1$ norm $\lnorm{\cdot}{1}$ & $\A_{\C^d}$ & trace (nuclear) norm $\norm{\cdot}{1}$\\
$\A^\circ_{\C^1}$ & element-wise max norm $\lnorm{\cdot}{\infty}$ & $\A^\circ_{\C^d}$ & operator norm $\norm{\cdot}{\infty}$\\
$\A^\circ_{\C^1_+}$ & maximum of diagonal elements and $0$ & $\A^\circ_{\C^d_+}$ & max. of eigenvalues and $0$\\
$\A^\circ_{\smash{{\C^1_+} \cup (-{\C^1_+})}}$ & maximum of absolute diagonal elements $\;\;$ & $\A^\circ_{\smash{{\C^d_+} \cup (-{\C^d_+})}}$ & numerical radius \cite{horn_2012}\vspace*{3pt}\\
$\Rs{\C^1_+}$ & standard robustness of coherence & $\Rs{\C^d_+}$ & trace of positive part $-1$ \\
$\Rg{\C^1_+}$ & generalised robustness of coherence \cite{piani_2016} & $\Rg{\C^d_+}$ & trace of positive part $-1$\\
$\A^\cup_{\C^1_+} - 1$ & coherence concurrence \cite{yuan_2015,qi_2017} & $\A^\cup_{\C^d_+} - 1$ & trace norm $-1$\\
$1 - \A^{\circ \cap}_{\C^1_+}$ & geometric measure of coherence \cite{streltsov_2015}&$1 - \A^{\circ \cap}_{\C^d_+}$ & ---\\\bottomrule
\end{tabular}
\end{adjustbox}\vspace*{5pt}
where by ``---'' we denote functions which, to our knowledge, have not been previously defined in the literature explicitly or do not correspond to easily characterisable quantities, although they can of course be defined in a similar manner. By the trace of the positive part of a Hermitian matrix $M$ we understand the quantity $\Tr(\Pi_+ M)$ with $\Pi_+$ denoting the orthogonal projection onto the subspace spanned by the non-negative eigenvalues of $M$. Further, we recall that $\A_{\C^d_+\cup(-\C^d_+)}$ corresponds to the trace norm itself for all Hermitian matrices \cite{hartkamper_1974}.

Note also that the set $\C^1_+$ only spans the set of diagonal matrices in the given basis, which means that the measure $\Rs{\C^1_+}$ is not a valid measure of $2$-coherence as it is infinite for $\rho \notin \C^1_+$.

\subsubsection{\texorpdfstring{$1 \leq k \leq d$}{1 <= k <= d}}

We will now consider the extension of the atomic gauge formalism to arbitrary $k$. Several of the measures in this resource theory have been defined in the literature already: they are the geometric measure of $k$-coherence \cite{regula_2018-2} and the generalised robustness $\Rg\Ckp$ \cite{ringbauer_2017}. All of the other measures that we have considered in Sec. \ref{sec:quantum} can be defined analogously. We will consider some explicit examples.

The sets $\Ckp$ for $k \geq 2$ have been shown to span the whole space $\HH$ \cite{ringbauer_2017}, and so $\Rs{\C^k_+}$ can be used as a quantifier for $k\geq 2$. The other considered measures $\A_{\I^k}$, $\A_{\C^k}$, $\A^\cup_\Ckp$, and $\Rg\Ckp$ have no domain problems because $\sspan(\I^k) = \CC^d \;\forall k$.

The atomic vector norm on the set $\I^k$, called the $k$-support norm $\norm{\cdot}{(k)}$, was introduced in \cite{argyriou_2012} in the context of machine learning and optimisation. The norm is in fact exactly computable for any $k$.

\begin{boxed}{white}
\makeatletter
\textbf{Definition} (Argyriou et al. \cite{argyriou_2012}). Let $\mathscr{P}_k$ be the set of all possible subsets of $\{1, \ldots, d\}$ of at most $k$ elements, and let $\esupp(\ket{x}) \coloneqq \lset i \bar x_i \neq 0\rset$ where $x_i$ is the $i$th coefficient of $\ket{x}$ in the given basis. The
\textbf{$k$-support norm}$^{\ref{foot}}$ is defined as the atomic norm for the set $\I^k$ and is given by
\begin{align}
\A_{\I^k}(\ket{x}) = \norm{\ket{x}}{(k)} &\coloneqq \inf \lset \sum_{I \in \mathscr{P}_k} \lnorm{\ket{v_I}}{2} \bar \esupp(\ket{v_I}) \subseteq I,\; \sum_{I \in \mathscr{P}_k} \ket{v_I} = \ket{x} \rset\nonumber\\
&= \sqrt{ \sum_{i=1}^{k-r-1} |x^\downarrow_i| ^2 + \frac{1}{r+1} \left(\sum_{i=k-r}^{d} |x^\downarrow_i| \right)^2 }\label{eq:ksupport_norm}
\end{align}
where $x^\downarrow$ denotes the coefficients of $\ket{x}$ sorted in non-increasing order by magnitude ($|x^\downarrow_1| \geq |x^\downarrow_2| \geq \ldots$), and $r$ is the unique integer in $\{1, \ldots, k-1\}$ satisfying
\begin{equation}
|x^\downarrow_{k-r-1}| > \frac{1}{r+1} \sum_{i=k-r}^{d} |x^\downarrow_{i}| \geq |x^\downarrow_{k-r}|
\end{equation}
or $0$ if no such integer exists.
\end{boxed}
\footnotetext[1]{\label{foot}The terminology of ``$k$-support'' comes from the fact that the set of indices corresponding to non-zero coefficients of a matrix is frequently referred to as ``support'' in statistical learning and optimisation literature \cite{bach_2012-1}. Since this differs from the usual definition of ``support'' in quantum information, where it is used to mean the orthogonal complement of the kernel of an operator \cite{wilde_2017}, to avoid confusion we use the notation of $\esupp$ to refer to the ``element-wise support''.}
It is explicit from Eq. \eqref{eq:ksupport_norm} that the $k$-support norm interpolates between the $\ell_1$ and $\ell_2$ norms for vectors --- indeed, we have that $\norm{\cdot}{(1)} = \lnorm{\cdot}{1}$ and $\norm{\cdot}{(d)} = \lnorm{\cdot}{2}$.

The polar (dual norm) of the $k$-support norm is given by
\begin{equation}
\norm{\ket{x}}{(k)}^\circ = \sqrt{\sum_{i=1}^{k} |x^\downarrow_i|^2},
\end{equation} 
that is, the $\ell_2$ norm of the $k$ largest coefficients. In general, we have that $\lnorm{\ket{x}}{1} \geq \norm{\ket{x}}{(k)} \geq \lnorm{\ket{x}}{2} \geq \norm{\ket{x}}{(k)}^\circ \geq \lnorm{\ket{x}}{\infty}$.

We remark that an alternative way to derive the exact formula for the $k$-support norm is to start with the dual norm $\norm{\cdot}{(k)}^\circ$, which is easier to compute explicitly \cite{regula_2018-2}, and apply the duality result from \cite{mudholkar_1985} as has been done in \cite{johnston_2015} for entanglement.

The atomic norm $\A_{\Ck}$ on the set $\Ck$ is the \textbf{(k,k)-trace norm}, defined as \cite{richard_2014,richard_2014-1}:
\begin{align}
  \norm{\rho}{(k,k)} &= \inf \lset \sum_i \norm{\ket{a_i}}{(k)} \norm{\ket{b_i}}{(k)} \bar \rho = \sum_i \ketbra{a_i}{b_i} \rset\nonumber\\
  &= \inf \lset \sum_{\substack{I \in \mathscr{P}_k\\J \in \mathscr{P}_k}} \norm{Z_{I,J}}{1} \bar \esupp\left(Z_{I,J}\right) \subseteq I \times J,\; \rho = \sum_{I,J} Z_{I,J} \rset\\
  \norm{\rho}{(k,k)}^\circ &= \sup \lset \cbraket{a | \rho | b} \bar \ket{a},\ket{b} \in \V^k \rset
\end{align}
where $\esupp(M)$ is defined for a matrix with elements $M_{ij}$ as $\esupp(M) = \lset (i,j) \bar M_{ij} \neq 0 \rset$. This norm interpolates between the entrywise $\ell_1$ norm for $\norm{\cdot}{(1,1)}$ and the standard trace norm for $\norm{\cdot}{(d,d)}$. The $(k,k)$-trace norm is therefore a natural generalisation of the $\ell_1$-norm of coherence, a fundamental measure of quantum coherence \cite{baumgratz_2014,rana_2017}, to the formalism of $k$-coherence.

While $\norm{\cdot}{(k,k)}$ is in general NP-hard to compute exactly, a numerical algorithm to approximate it was introduced in \cite{richard_2014}. Note also that the generalised robustness of $k$-coherence $\Rg{\C^k_+}$ was shown to be computable with a semidefinite program (SDP) \cite{ringbauer_2017}, and it follows easily that the same property holds for the standard robustness $\Rs{\C^k_+}$. We stress the general relation $\norm{\rho}{(k,k)} - 1 \geq \Rg{\Ck} (\rho)$ with equality for pure states.

The convex roof extension of the $k$-support norm $\A^\cup_\Ckp$ is the natural convex roof measure in this formalism. By Theorem \ref{thm:monotone_conv}, both the convex roof--extended $k$-support norm as well as the $(k,k)$-trace norm are strongly monotonic under $k$-incoherent operations \cite{ringbauer_2017}. We remark that another convex roof--based measure was defined in \cite{chin_2017}, based on an entanglement monotone called $k$-concurrence; it is not a generalisation of the $\ell_1$ norm in the present framework, and we will compare the measures quantitatively in the next section (as the pure-state coherence and entanglement measures will turn out to be equivalent).


\subsection{Bipartite entanglement}\label{sec:bipartite_entanglement}

Let $\ket{\psi} \in \CC^{d_A} \otimes \CC^{d_B}$ denote a bipartite pure state shared between two parties $A$ and $B$, with the dimensions of the corresponding spaces denoted as $d_A$ and $d_B$. A pure state is called separable or a product state (in the $A|B$ bipartition) if it can be written as $\ket{\psi} = \ket{\psi_A} \otimes \ket{\psi_B}$, and entangled otherwise. We denote the set of all product state vectors as $\V$, and we define $\Sp$ to be the set of all separable density matrices defined through the convex hull. The \textit{Schmidt rank} of a state $\ket\psi$ is defined as
\begin{equation}
  \SR(\ket\psi) = \min \left\{ r \in \mathbb{N} \,\left|\; \ket{\psi} = \sum_{i=1}^{r} \lambda_i \ket{c_i},\, \ket{v_i} \in \V \right.\right\}.
\end{equation}
Then $\ket{\psi}$ is unentangled if and only if $\SR\left(\ket{\psi}\right) = 1$, and in general we have $1 \leq \SR(\ket\psi) \leq \min(d_A, d_B)$. A crucial property of bipartite entanglement is that every state can be expressed in the so-called Schmidt decomposition as
\begin{equation}
  \ket\psi = \sum_{i=1}^{\SR(\ket\psi)} \lambda_i \ket{a_i} \otimes \ket{b_i} 
\end{equation}
where $\{\ket{a_i}\}$ and $\{\ket{b_i}\}$ form orthonormal bases for $\CC^{d_A}$ and $\CC^{d_B}$, respectively, and the terms $\lambda_i$ are called \textit{Schmidt coefficients}. 

As previously, we now want to consider situations where not all entangled states are resourceful; that is, for our particular task, a state with at least a given Schmidt rank $k$ is necessary, and we would like to quantify the entanglement corresponding to a particular Schmidt rank \cite{terhal_2000,chruscinski_2009,johnston_2010,sentis_2016}. We define the Schmidt vector $\lambda(\ket\psi)$ of a state $\ket\psi$ to be the vector consisting of its Schmidt coefficients including the zero terms, so that it is always $\min(d_A,d_B)$-dimensional. Since the Schmidt rank of a state corresponds to the cardinality of the Schmidt vector just as the coherence rank corresponds to the cardinality of the state vector in the given basis, in the case of pure states, the resource theory of bipartite entanglement of Schmidt rank $k+1$ can be seen to correspond to the resource theory of $k+1$-coherence applied to the Schmidt vector. Indeed, this relation between the two resource theories has been used to relate the corresponding quantifiers \cite{chen_2016,zhu_2017,zhu_2017-1}.

We now define the relevant sets of free pure state vectors and free density matrices as, respectively:
\begin{equation}\begin{aligned}
  \V^k &= \lset \ket{\psi} \bar \SR\left(\ket{\psi}\right) \leq k, \braket{\psi|\psi} = 1 \rset\\
  \S^k_+ &= \conv\lset \proj\psi \bar \ket\psi \in \V^k \rset
\end{aligned}\end{equation}
and define the corresponding set
\begin{equation}
  \S^k = \conv\lset \ket\psi\bra\phi \bar \ket\psi \in \V^k,\;\ket\phi \in \V^k \rset.
\end{equation}
$\S^1_+$ is then the set of separable states, and $\S^d_+ = \DD$. A mixed state $\rho \in \S^k_+$ is said to have Schmidt number (at most) $k$ \cite{terhal_2000}. The dual cone ${\mathcal{S}^k_{+}\hspace{-0.0ex}\raisebox{0.2ex}{\*}}$ defines the set of the so-called $k$-block positive operators \cite{skowronek_2009}.

Note that for any $k$, we have that $\sspan(\V^k)=\CC^d$ and $\sspan(\S^k_+) = \HH$ \cite{zyczkowski_1998}, which in particular means that all of the considered functions are finite for any density matrix, and that the relevant symmetric gauges ($\A_{\V^k}, \A_{\S^k_+ \cup \left(-\S^k_+\right)}, \A_{\S^k}$) all define valid norms.

The set of operations of interest in the theory of entanglement --- \locc{} --- is a strict subset of both the resource non-generating (separability-preserving) and the stochastically resource non-generating (separable) operations \cite{chitambar_2014}, which by the results of Sec. \ref{subsec:resource_quantifiers} means that all of the considered measures are strong monotones under \locc. We stress that the measures introduced herein are monotonic not only under free operations which do not generate entanglement, but also under a larger set of operations which do not generate entanglement of Schmidt rank $k+1$. 

\subsubsection{\texorpdfstring{$k=1$}{k=1}}

For $k=1$, we obtain many familiar measures and quantifiers in this formalism:

\begin{center}
\begin{tabular}{@{} *2l @{}}  \toprule
Gauge & Also known as\\\midrule
$\A_{\V^1}$ & sum of Schmidt coefficients\\
$\A^\circ_{\V^1}$ & largest Schmidt coefficient\\
$\A_{\S^1}$ & greatest cross norm (projective tensor norm) \cite{rudolph_2001,rudolph_2005}\\
$\A^\circ_{\S^1}$ & Schmidt operator norm (injective tensor norm) \cite{johnston_2010}\\
$\A^\circ_{\S^1_+}$ & ---\\
$\A^\circ_{\smash{{\S^1_+} \cup (-{\S^1_+})}}$ & product numerical radius \cite{gawron_2010}\vspace*{3pt}\\
$\Rs{\S^1_+}$ & robustness of entanglement \cite{vidal_1999}\\
$\Rg{\S^1_+}$ & generalised robustness of entanglement \cite{vidal_1999,steiner_2003}\\
$\A^\cup_{\S^1_+} - 1$ & ($2\times$) convex roof--extended negativity \cite{lee_2003}\\
$1-\A^{\circ \cap}_{\S^1_+}$ & geometric measure of entanglement \cite{vidal_1999-1,wei_2003}\\\bottomrule
\end{tabular}
\end{center}

The vector gauge $\A_{\V^1}$ is the $\ell_1$ norm of the Schmidt vector, corresponding to the sum of Schmidt coefficients and constituting a convex relaxation of the Schmidt rank of a pure state.

The negativity of a state is defined as $N(\rho) = \frac{1}{2}\left(\norm{\rho^{T_B}}{1} - 1\right)$ \cite{zyczkowski_1998,vidal_2002}, where $\rho^{T_B}$ denotes the partial transpose. The negativity of pure states is precisely \cite{vidal_2002}
\begin{equation}\begin{aligned}
N(\proj\psi) &= \sum_{j<k} \lambda_j \lambda_k =\frac{\A_{\V^1}(\ket\psi)^2 - 1}{2},
\end{aligned}\end{equation}
which means that the convex roof--based quantifier $\A^\cup_{\S^1}$ is (twice) the \textit{convex roof--extended negativity}. This function was proposed as an alternative generalisation of concurrence to systems beyond two qubits \cite{lee_2003}, and it was further suggested as the measure most suitable to characterise the so-called monogamy relations of entanglement between qudits \cite{ou_2007,kim_2009,regula_2014,choi_2015,tian_2016,regula_2016-1}, which the concurrence fails to satisfy. Here we also see that the convex roof--extended negativity arises as the natural gauge-based generalisation of the concurrence. The faithfulness and strong monotonicity of $\A^\cup_{\S^1}$ under separable operations follow from Theorem \ref{thm:monotone_conv}. 

It has been pointed out that the quantity $2 N(\rho)+1$ can be used to lower bound the Schmidt number of a given quantum state \cite{eltschka_2013}. The atomic gauge function formalism provides a geometric intuition and justification for this statement, in the sense that for pure states $\A_{\V^1}(\ket\psi) = 2 N(\proj\psi) + 1$ is exactly a natural convex relaxation of the Schmidt rank, and the functions $\A^\cup_{\S^1}(\rho)$, $\A_{\S^1}(\rho)$ as well as the robustness measures all constitute convex lower bounds to the Schmidt number of a state. The negativity itself then constitues a lower bound to the gauges (see Sec. \ref{sec:ppt} below).

The atomic norm $\A_{\S^1}$ is the greatest cross norm, a quantifier introduced in the context of entanglement by Rudolph \cite{rudolph_2000,rudolph_2001}. It can alternatively be written as
\begin{equation}
  \A_{\S^1}(\rho) = \inf \lset \sum_i \norm{X^{(i)}_A}{1} \norm{X^{(i)}_B}{1} \bar \rho = \sum_i X^{(i)}_A \otimes X^{(i)}_B,\; X^{(i)}_A \in \mathbb{C}^{d_A \times d_A},\; X^{(i)}_B \in \mathbb{C}^{d_B \times d_B} \rset.
\end{equation}

\subsubsection{\texorpdfstring{$k\geq1$}{k>=1}}

The generalisation of many of the above quantities to the set of bipartite states with a given Schmidt number $k$ was considered by Johnston and Kribs in \cite{johnston_2010,johnston_2011,johnston_2015}, and the generalised robustness in \cite{clarisse_2006}. Similarly to the case of coherence, we obtain a hierarchy of quantifiers, each corresponding to a different level of bipartite entanglement. 

\begin{figure}[t]
\centering
\begin{adjustbox}{center}
\begin{subfigure}{.33\textwidth}
  \centering
  \includegraphics[width=5.2cm]{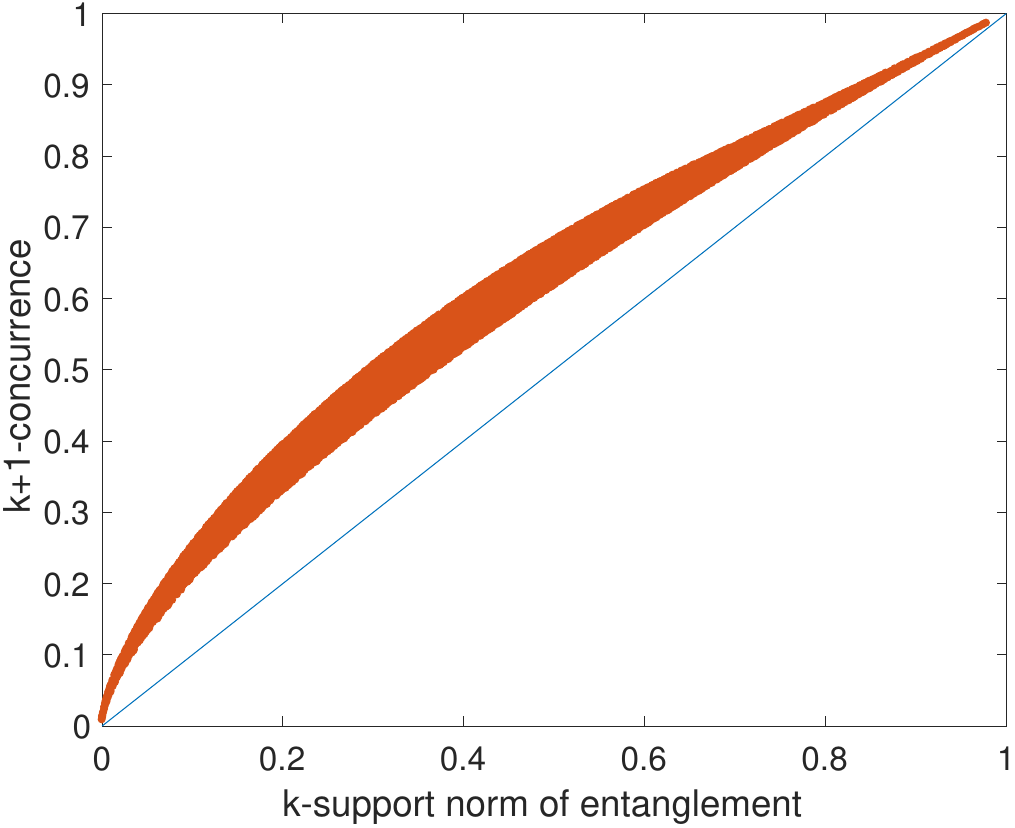}
  \caption{$d_A = d_B =3,\; k=2$}
  \label{fig:sub1}
\end{subfigure}\hspace{3pt}
\begin{subfigure}{.33\textwidth}
  \centering
  \includegraphics[width=5.2cm]{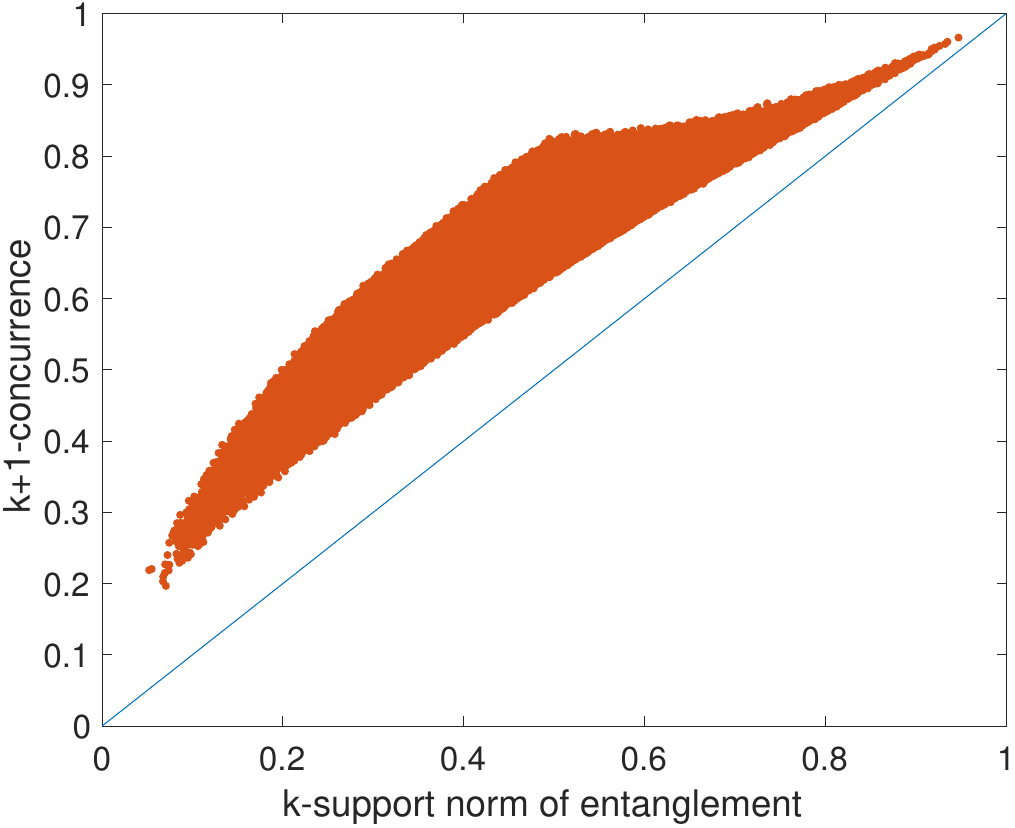}
  \caption{$d_A = d_B = 4,\; k=2$}
  \label{fig:sub2}
\end{subfigure}\hspace{3pt}
\begin{subfigure}{.33\textwidth}
  \centering
  \includegraphics[width=5.2cm]{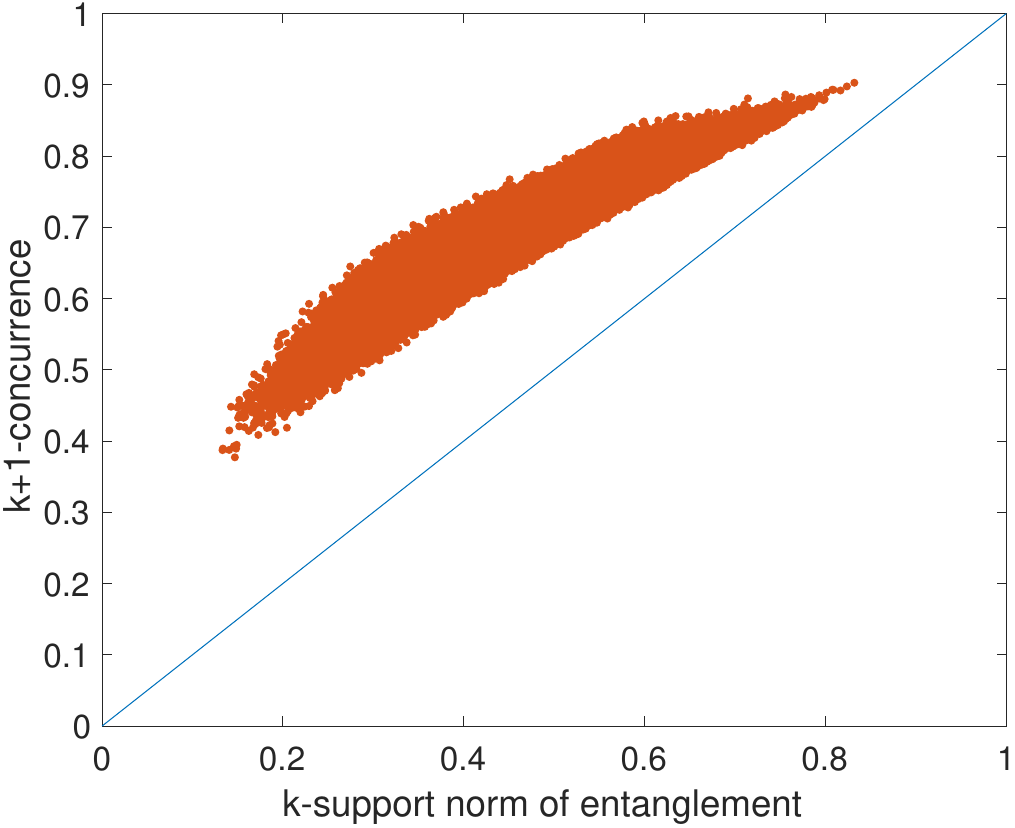}
  \caption{$d_A = d_B = 6,\; k=3$}
  \label{fig:sub3}
\end{subfigure}
\end{adjustbox}
\captionsetup{width=.97\linewidth,justification=raggedright}
\caption{\normalsize Comparison of the different Schmidt rank $k$ entanglement measures for $10^6$ randomly generated pure states (uniformly distributed with respect to the Haar measure). On the $x$ axis is the quantity $\A_{\V^k}(\ket\psi)^2-1 = \norm{\lambda(\ket\psi)}{(k)}^2 - 1$ (with a normalisation factor $\frac{k}{d_A-k}$), and on the $y$ axis is the $k+1$-concurrence monotone from \cite{gour_2005}.}
\label{fig:ksupport}
\end{figure}

The quantity $\A_{\V^k}$ is nothing but the $k$-support norm of the Schmidt vector, $\norm{\lambda(\cdot)}{(k)}$. This gives a natural generalisation of the convex roof--extended negativity to a measure of Schmidt rank $k$ entanglement:
\begin{equation}
  \A^\cup_{\S^k}(\rho) = \inf \lset \sum_i p_i \norm{\lambda\left(\ket{\psi_i}\right)}{(k)}^2 \bar \rho = \sum_i p_i \proj{\psi_i},\,p_i\in\RR_+,\,\sum_i p_i = 1 \rset.
\end{equation}
Another common convex roof--based measure of Schmidt rank $k$ entanglement is Gour's $k+1$-concurrence \cite{gour_2005} --- a comparison between the values of $\A_{\V^k}$ and the $k+1$-concurrence on pure states can be found in Fig. \ref{fig:ksupport}.

The norm $\A_{\S^k}$ generalises the greatest cross norm to a faithful quantifier of entanglement of a given Schmidt rank \cite{johnston_2015}. One can also express $\A_{\S^k}$ as the nuclear norm
\begin{equation}
  \A_{\S^k} (\rho) = \inf \lset \sum_i \norm{\lambda\left(\ket{x_i}\right)}{(k)}\norm{\lambda\left(\ket{y_i}\right)}{(k)} \bar \rho = \sum_i \ketbra{x_i}{y_i} \rset.
\end{equation}
We remark that, just as in the case of coherence, the robustness $\Rg{\S^k_+}$ provides a tight lower bound for this norm. Further, we establish that $\Rg{\S^k_+}$ reduces on pure states to the $k$-support norm of entanglement, thus generalising the known relation between the robustness of entanglement and negativity \cite{vidal_1999}. Note also that the polar gauge $\A^\circ_{\S^k}$ can be computed exactly in small dimensions, and in general bounded by semidefinite programs \cite{johnston_2011}, allowing for an efficient characterisation of $k$-block positive operators \cite{johnston_2010}. 

The quantity $1-\A^{\circ \cap}_{\S^k_+}$ provides a generalisation of the geometric measure of entanglement, and in fact corresponds to a family of convex roof--based monotones introduced by Vidal \cite{vidal_1999-1}.

\subsubsection{Remarks about the resource theory of negative partial transpose}\label{sec:ppt}

Letting $T_B$ denote the transpose map on $\CC^{d_B}$, the partial transpose of $\rho$ is given by $\rho^{T_B} = (\mathds{1}_A \otimes T_B)(\rho)$. The set of states with positive partial transpose (\PPT) is then $\pptp = \lset \rho \in \DD \bar \rho^{T_B} \in \HH_+ \rset$. It is well-known that $\S^1_+ \subseteq \pptp$ with equality iff $d_A d_B \leq 6$ \cite{horodecki_1996,horodecki_1997}, and that the separable operations (stochastically resource non-generating in the resource theory of bipartite entanglement) are a subset of the so-called \PPT{} operations \cite{horodecki_1997,chitambar_2014}, which themselves belong to the set of resource non-generating operations in the resource theory of negative partial transpose. What this means in particular is that, by considering the resource theory in which $\pptp$ is the set of free states, any monotone under \PPT{} operations is an entanglement monotone under separable operations --- including the quantities $\Rs{\pptp}$ and $\Rg{\pptp}$.

A different choice of the reference set of free states is often made, taking instead the set of unit trace \textit{Hermitian} matrices with a positive partial transpose, $\ppt = \lset X \in \HH \bar \<\mathds{1},X\> = 1,\right.$ $\left.X^{T_B} \in \HH_+ \rset$. The gauge function formalism straightforwardly applies to such cases as well --- indeed, the negativity of any mixed state, defined as $N(\rho) = \frac{1}{2}\left(\norm{\rho^{T_B}}{1}-1\right)$, can be written simply as $N(\rho) = \Rs\ppt(\rho)$ \cite{vidal_2002,plenio_2007}. The function $\Rs{\pptp}$ was in fact mentioned in the original work of Vidal and Werner \cite{vidal_2002} as an alternative to negativity, leading to the relation $\Rs{S^1_+}(\rho)\geq \Rs{\pptp}(\rho) \geq N(\rho)$ by the set inclusion of their unit balls. More recently, other gauge functions in the resource theory of negative partial transpose have been employed to provide bounds on distillable entanglement \cite{wang_2016} and characterise asymptotic entanglement manipulation \cite{wang_2017-3}.

These considerations generalise also to multipartite entanglement in the form of quantifiers such as the genuine multipartite negativity \cite{jungnitsch_2011,hofmann_2014}, as we will consider explicitly in the next section.

\subsection{\texorpdfstring{$k$}{k}-partite entanglement}

Let us now consider a system consisting of $n\geq2$ parties, $\ket\psi \in \CC^{d_1} \otimes \cdots \otimes \CC^{d_n}$. Similarly to the hierarchy of the Schmidt rank in the case of bipartite entanglement, one can define a hierarchy of multipartite entanglement. We will take $\P^{k}$ with $1 \leq k \leq n$ to be the set of pure states which are $k$-producible --- that is, they can be expressed as \cite{guhne_2005}
\begin{equation}
  \ket\psi = \ket{\phi_1} \otimes \ket{\phi_2} \otimes \cdot \otimes \ket{\phi_m}
\end{equation}
where each of the $m$ states $\ket{\phi_i}$ consists of at most $k$ parties. Since such states are at most $k$-partite entangled, they are precisely the free states in the resource theory of $k+1$-partite entanglement. The convex hull of the corresponding set of matrices $\M^k_+ = \conv \lset \proj{a} \bar \ket{a} \in \P^k \rset$ then defines the set of $k$-producible mixed states, and we analogously define $\M^k = \conv \lset \ketbra{a}{b} \bar \ket{a}, \ket{b} \in \P^k \rset$. Note that $\M^1_+$ is the set of fully separable states and $\M^n_+ = \DD$.

Unlike the theory of bipartite entanglement, our understanding of the theory of multipartite entanglement is still very limited \cite{horodecki_2009,walter_2016}. While one can straightforwardly generalise the quantities defined previously for bipartite entanglement --- for instance, as the robustness of $k$-partite entanglement \cite{cavalcanti_2005}, geometric measure of $k$-partite entanglement \cite{blasone_2008}, and product numerical radius \cite{puchala_2011} --- their quantification and characterisation is in general a much more difficult task.

In the case of $k=1$, the corresponding norm $\A_{\M^1}$ generalises the greatest cross norm \cite{rudolph_2001}. This quantity along with its dual norm $\A^\circ_{\M^1}$ have been investigated in detail in \cite{arveson_2009}, including the infinite-dimensional case. The norms $\A_{\P^1}$ and $\A_{\M^1}$ were in fact shown to correspond to the same concept --- projective tensor norm --- but defined on different spaces, the Hilbert space $\mathcal{H} \cong \CC^{d_1} \otimes \cdots \otimes \CC^{d_n}$ in the case of $\A_{\P^1}$, and the Hilbert space of trace-class bounded linear operators on $\mathcal{H}$ in the case of $\A_{\M^1}$. Explicitly, one can write
\begin{align}
  \A_{\P^1} (\ket\psi) &= \inf \lset \sum_i \lnorm{\ket{a^{(i)}_1}}{2}\lnorm{\ket{a^{(i)}_2}}{2}\cdots\lnorm{\ket{a^{(i)}_n}}{2} \bar \ket\psi = \sum_i \ket{a^{(i)}_1} \otimes \ket{a^{(i)}_2} \otimes \cdots \otimes \ket{a^{(i)}_n},\right.\nonumber\\
  &\hphantom{= \inf \lset \sum_i \lnorm{\ket{a^{(i)}_1}}{2}\lnorm{\ket{a^{(i)}_2}}{2}\cdots\lnorm{\ket{a^{(i)}_n}}{2} \bar\right.\;}\ket{a^{(i)}_j} \in \CC^{d_j} \Bigg\}\\
  \A_{\M^1} (\rho) &= \inf \lset \sum_i \norm{X^{(i)}_1}{1} \norm{X^{(i)}_2}{1} \cdots \norm{X^{(i)}_n}{1} \bar \rho = \sum_i X^{(i)}_1 \otimes X^{(i)}_2 \otimes \cdots \otimes X^{(i)}_n,\right.\nonumber\\
  &\hphantom{= \inf \lset \sum_i \norm{X^{(i)}_1}{1} \norm{X^{(i)}_2}{1} \cdots \norm{X^{(i)}_n}{1} \bar \right.\;} X^{(i)}_j \in \CC^{d_j\times d_j} \Bigg\}\label{eq:m1norm}
\end{align}
following \cite{rudolph_2001,arveson_2009}, where $d_j$ denotes the dimension of the Hilbert space of the $j$th system. Note that the base norm $\A_{\M^1_+ \cup (-\M^1_+)}$ can be expressed by replacing $X^{(i)}_j \in \CC^{d_j\times d_j}$ with $X^{(i)}_j \in \HH^{d_j\times d_j}$ in Eq. \eqref{eq:m1norm} \cite{rudolph_2005}.

Generalising this norm to the fine-grained classification of $k$-partite entanglement, one obtains a hierarchy of norms akin to the one defined for Schmidt rank $k$ entanglement. Explicitly, we have:
\begin{equation}\begin{aligned}\label{eq:kpart_entanglement_norms}
  \A_{\M^k} (\rho) &= \inf \lset \sum_i c_i \bar \rho = \sum_i c_i \ketbra{v_i}{w_i},\; \ket{v_i}, \ket{w_i} \in \P^k \rset\\
  &= \inf \lset \sum_i \A_{\P^k}(\ket{x_i}) \A_{\P^k}(\ket{y_i}) \bar \rho = \sum_i \ketbra{x_i}{y_i} \rset\\
  \A^\circ_{\M^k} (\rho) &= \sup \lset \cbraket{a | \rho | b} \bar \ket{a}, \ket{b} \in \P^k\rset
\end{aligned}\end{equation}
and analogously for the other definitions. The computation of these quantities is of course not easy, especially without being able to rely on tools such as the bipartite Schmidt decomposition, although let us remark that generalisations of the Schmidt decomposition have been proposed \cite{carteret_2000,acin_2000,sokoli_2014} and one of them in particular has been related to the quantification of the $\A_{\M^1}$ norm for some special cases of states \cite{sokoli_2014}.

A case of particular interest is the so-called \textit{genuine} $n$-partite entanglement, corresponding to states which are not separable across any bipartition; in other words, the resource theory whose free states are constituted by the $(n-1)$-producible states $\M^{n-1}_+$. This case is particularly simple to consider for pure states, where it suffices to minimise the bipartite atomic gauge over all possible $2^{n-1}-1$ bipartitions of the system:
\begin{equation}
  \A_{\P^{n-1}} (\ket\psi) = \;\min_{\mathclap{\substack{\text{bipartition}\\(A|B)}}}\enskip \A^{(A|B)}_{\V^1} \,(\ket\psi) 
\end{equation}
with $\A^{(A|B)}_{\V^1}$ denoting the atomic gauge $\A_{\V^1}$ introduced in Sec. \ref{sec:bipartite_entanglement} computed in the given bipartition $(A|B)$. Since we have shown $\A^2_{\V^1}-1$ to be equal to twice the bipartite negativity, it means that $\A^2_{\P^{n-1}}-1$ is in fact equal to twice the so-called renormalised genuine multiparticle negativity (\GMN) \cite{hofmann_2014}. The convex roof extension of $\A^2_{\P^{n-1}}-1$ then generalises the convex roof--extended negativity to a measure of genuine multipartite entanglement.

The \GMN{} itself, computable with a semidefinite program \cite{hofmann_2014}, can be used to provide lower bounds for the measures. Furthermore, we have that the quantifiers such as the generalised robustness of genuine multipartite entanglement are equal to twice the \GMN{} on pure states, generalising the known relation of the negativity and robustness for $n=2$ \cite{vidal_2002,steiner_2003} and allowing for an efficient quantification of genuine multipartite entanglement for pure states.

The properties of the quantities introduced above such as faithfulness, strong monotonicity under free operations (which includes \locc{} as a subset), quantitative bounds, and their relation to witnesses of $k$-partite entanglement all follow from the results of this work. We remark that since for each $k$, $\P^k$ and $\M^k_+$ span the whole space, the considered measures all have full domain and $\A_{\P^k}$, $\A_{\M^k}$, and $\A_{\M^k_+ \cup (-\M^k_+)}$ all define valid norms.

\subsection{Magic states}\label{sec:magic}

As the final example of the application of the atomic gauge formalism, we briefly consider the resource theory of magic states, recently characterised in \cite{veitch_2012,veitch_2014,howard_2014,delfosse_2015,bravyi_2016,raussendorf_2017,howard_2017,ahmadi_2017}. Here, the free pure state projectors are the so-called stabiliser states, consisting of the eigenvectors of Heisenberg-Weyl operators, and states outside of their convex hull are called \textit{magic states}.
 Since there is always a finite number of pure stabiliser states (although it scales superexponentially with the dimension of the Hilbert space \cite{gross_2006}), the set of free density matrices is given by
 \begin{equation}
   \W_+ = \conv \left\{ \proj{v_1}, \ldots, \proj{v_k} \right\}
\end{equation}
and forms a polytope in the space of Hermitian matrices, which means that the quantification of many measures in the atomic gauge formalism will reduce to solving a linear or semidefinite programming problem \cite{howard_2017,ahmadi_2017}. We will make this explicit by expressing the optimisation problems involved in simplified forms.

To begin, let us define the non-balanced set $\T' = \left\{ \ket{v_1}, \ldots, \ket{v_k} \right\}$, where $\{\ket{v_i}\}$ are the pure stabiliser states with any complex phase. Since we require the set of free states to be balanced, the set of interest $\T \subset \CC^d$ is given by $\T'$ symmetrised around the origin, which we can achieve by replacing real coefficients with complex ones in the definitions of the gauge functions. Precisely, the atomic gauge function corresponding to this set can be obtained by considering the $d \times k$ matrix whose columns correspond to the free states, i.e. $T = \big(\ket{v_1}, \ldots, \ket{v_k}\big)$, which then gives
\begin{equation}\begin{aligned}
  \A_\T(\ket\psi) &= \min \lset \sum_i c_i \bar \ket\psi = \sum_i c_i \ket{w_i},\; \ket{w_i} \in \T,\; c_i \in \RR_+ \rset\\
  &= \min \lset \sum_i |c_i| \bar \ket\psi = \sum_i c_i \ket{v_i},\; \ket{v_i} \in \T',\;c_i \in \CC \rset\\
  &= \min_{x \in \CC^k} \lset \lnorm{x}{1} \bar Tx = \ket\psi \rset
\end{aligned}\end{equation}
where we do not use the Dirac notation for $x \in \CC^k$ to differentiate the two different spaces $\CC^k$ and $\CC^d$. This quantity can be thought of as a convex relaxation of the stabiliser rank, quantifying the least number of stabiliser states one needs to superpose to express the state $\ket\psi$ \cite{bravyi_2016}.

Defining the set $\W = \lset \ketbra{w_i}{w_j} \bar \ket{w_i}, \ket{w_j} \in \T \rset$, it is straightforward to see from the above characterisation that the nuclear gauge of the set $\W$ is given by
\begin{equation}
  \A_\W (\rho) = \min_{X \in \CC^{k \times k}} \lset \lnorm{X}{1} \bar T X T^\dagger = \rho \rset.
\end{equation}
By the discussion in Sec. \ref{subsec:resource_quantifiers}, this defines a faithful monotone of magic. Since the set $\T$ spans the space $\CC^d$, we have that $\A_\W$ is a valid norm for all complex matrices. Again in a similar way, we get the atomic gauges corresponding to the robustness measures as
\begin{align}
  2\Rs{\W_+} (\rho) + 1 = \A_{\W_+\cup(-\W_+)} (\rho) &= \min_{x \in \RR^k} \lset \lnorm{x}{1} \bar T \diag(x) T^\dagger = \rho \rset\\
   \Rg{\W_+} (\rho) + 1 &= \min_{x \in \RR^k_+} \lset \lnorm{x}{1} \bar T \diag(x) T^\dagger \cgeq \rho \rset.
\end{align}
Although no longer a linear program, the generalised robustness has the advantage that, once again, its quantification reduces to computing the gauge $\A_\T$ for pure states by Theorem \ref{prop:gen_robustness_pure_states}. An alternative characterisation of the above quantities in terms of generalised Bloch vectors can be obtained by following \cite{howard_2017}, where we remark that in \cite{howard_2017} the quantity $\A_{\W_+ \cup (-\W_+)}$ itself was referred to as the robustness of magic.

\begin{figure}[h!]
  \centering
  \makebox[\textwidth][c]{
  \begin{subfigure}{.49\textwidth}
  \includegraphics[width=7.46cm]{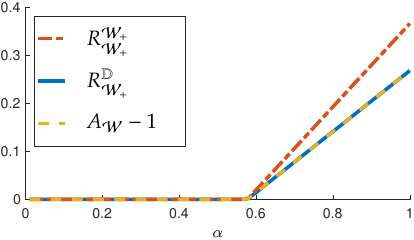}
\caption{}
\end{subfigure}\hspace{10pt}
\begin{subfigure}{.49\textwidth}
   \includegraphics[width=7.46cm]{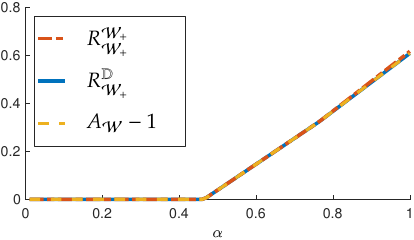}
\caption{}
\end{subfigure}
}
\captionsetup{width=.9\linewidth,justification=raggedright}
\begin{center}
  \caption{Quantitative comparison of the measures of magic for: (a) the one-qubit state $(1-\alpha) \frac{\mathds{1}}{2} + \alpha \proj{T}$, (b) the two-qubit state $(1-\alpha) \frac{\mathds{1}}{4} + \alpha \proj{T\otimes T}$.}
  \label{fig:magic}
  \end{center}
  \end{figure}

Note that the resource theory of magic states is an example of a resource theory where the standard robustness $\Rs{\W_+}$ is in general strictly larger than the other measures on pure states: as an explicit example, consider the one-qubit pure state $\proj{T} = \frac{1}{2}\left(\mathds{1} + \frac{\sigma_x + \sigma_y + \sigma_z}{\sqrt{3}}\right)$ \cite{bravyi_2005} with $\sigma_i$ being the Pauli operators. The standard robustness of this state can be computed exactly as $\Rs{\W_+}(\proj{T}) = \frac{1}{2}(\sqrt{3}-1)$ \cite{howard_2017}, while a similar calculation for the other measures yields
\begin{equation}
  \A_{\T}(\ket{T})^2-1 = 2 - \sqrt{3} \approx 0.268 < \Rs{\W_+} (\proj{T}) = \frac{\sqrt{3}-1}{2} \approx 0.366
\end{equation}
where we recall that $\Rg{\W_+}(\proj{T}) = \A_{\W}(\proj{T})-1 =\A_{\T}(\ket{T})^2-1$. However, the difference appears to become less pronounced with increasing dimension, with the two-qubit state $\ket{T \otimes T}$ having standard robustness of approx. $0.616$ and generalised robustness of approx. $0.607$. A comparison of the measures for a class of mixed states is plotted in Fig. \ref{fig:magic}.

The convex roof--based quantifier in this formalism can be defined analogously. By noting again that each $\ket{\psi_i} \in \T\*\*$ can be written as $\ket{\psi_i} = T y$ for some $y \in \CC^k$, we get that any $P \in \lset \proj{\psi_i} \bar \ket{\psi_i} \in \T\*\* \rset\*\*$ can be expressed as $P = \sum_{i=1}^r \proj{\psi_i} = T Y Y^\dagger T^\dagger$ for some $Y \in \CC^{k \times r}$, where we can take $r \leq \rank(P)^2$ by Carath\'{e}odory's theorem \cite{uhlmann_1998}. We can then write
\begin{equation}
  \A^\cup_{\W_+} (\rho) = \min_{Y \in \CC^{k\times r}} \lset \norm{Y}{\ell_1,\ell_2}^2 \bar T Y Y^\dagger T^\dagger = \rho \rset
\end{equation}
where $\norm{Y}{\ell_1,\ell_2}^2 = \sum_{j=1}^r\big(\sum_{i=1}^k |Y_{ij}|\big)^2$. As for other quantum resources, this problem is in general more difficult to solve than the other quantifiers.

We remark that a very similar formalism applies to any resource theory where the set of free states is finite. For example, the resource theory of quantum coherence ($k=1$) can be obtained by taking $T$ to consist of the vectors of the reference orthonormal basis, although care has to be taken about the effective domains of the measures as discussed before.

\section{Conclusions}

We have introduced a framework for the quantification of arbitrary convex quantum resources based on the atomic gauge functions of the corresponding sets of free states. We have shown that the formalism encompasses many commonly used measures and allows for a straightforward comparison and characterisation of the quantifiers. In addition to the measures explicitly introduced herein --- such as ones based on matrix norms, the convex roof, or the robustness measures --- we have shown that the framework can be applied to describe more general kinds of quantifiers, and we provided easily verifiable conditions guaranteeing that a given measure satisfies desirable properties such as faithfulness and strong monotonicity under the free operations of the resource theory. Further, we have explicitly applied the results to the resource theories of quantum coherence, entanglement, and magic states, establishing a detailed characterisation of many known monotones as well as introducing novel measures for the resources. 

The results presented here can be generalised in several ways. Firstly, note that the formalism of gauge functions allows for an application of the same concepts to infinite-dimensional spaces, as has already been done for entanglement in \cite{arveson_2009,shirokov_2010}. Secondly, while we focused on the application of our framework to quantum resource theories, it can of course be used for arbitrary convex sets, beyond quantum resources and quantum states. It would also be interesting to investigate in more detail the relation between quantifiers based on atomic gauges and ones based on distance measures \cite{vedral_1997,bengtsson_2007}, as well as quantifiers which are defined through algebraic properties but can nevertheless admit a geometric interpretation --- such as measures of entanglement obtained from polynomial invariants \cite{viehmann_2011,eltschka_2014-1,regula_2016,regula_2016-2}.

We hope that our investigation into the convex geometry of measures of quantum resources will contribute to a better understanding of the properties and interrelations between various resource quantifiers, as well as provide an accessible framework to define and characterise the measures of any given convex quantum resource, complementing the recent efforts to establish a unified mathematical description of convex resources \cite{sperling_2015,brandao_2015,gour_2017,girard_2014}.

\subsubsection*{Acknowledgements}
I am very grateful to Gerardo Adesso, Thomas R. Bromley, Marco Cianciaruso, Nathaniel Johnston, and Marco Piani for many discussions and suggestions about the preliminary drafts of this manuscript, as well as to Ludovico Lami, Earl T. Campbell, and Namit Anand for useful comments. This work has been supported by the European Research Council (ERC) under the Starting Grant GQCOP (Grant No.~637352).

\enlargethispage{\baselineskip}
\bibliographystyle{apsrev4-1}
\bibliography{main}

\end{document}